\documentclass[a4paper,11pt]{article}
\pdfoutput=1 

\usepackage{jinstpub} 
\usepackage{lineno}
\usepackage{verbatim} 
\usepackage{multirow}
\usepackage{amsmath}

\title{\boldmath The ATLAS Readout System for LHC Runs 2 and 3}

\author[a]{A. Borga,}
\author[h]{R. Blair,}
\author[b]{G.J. Crone,}
\author[c]{B. Green,}
\author[d]{A. Kugel,}
\author[e]{M. Joos,}
\author[h]{J. Love,}
\author[c,1]{J.G. Panduro Vazquez,\note{Corresponding author.}}
\author[e,f]{J. Schumacher,}
\author[c]{P. Teixeira-Dias,}
\author[e]{L. Tremblet,}
\author[e]{W. Vandelli,}
\author[a]{J.C. Vermeulen,}
\author[h]{O. Rifki,}
\author[e]{P. Werner,}
\author[g]{F.J. Wickens}

\affiliation[a]{Nikhef, National Institute for Subatomic Physics and University of Amsterdam, Amsterdam, The Netherlands}
\affiliation[b]{Department of Physics and Astronomy, University College London, London, United Kingdom}
\affiliation[c]{Department of Physics, Royal Holloway University of London, Surrey, United Kingdom}
\affiliation[d]{ZITI Institut f{\"u}r technische Informatik, Ruprecht-Karls-Universit{\"a}t Heidelberg, Mannheim,Germany}
\affiliation[e]{CERN, Geneva, Switzerland}
\affiliation[f]{Department of Computer Science, University of Paderborn, Paderborn, Germany}
\affiliation[g]{Particle Physics Department, Rutherford Appleton Laboratory, Didcot, United Kingdom}
\affiliation[h]{High Energy Physics Division, Argonne National Laboratory, Lemont, Illinois, United States of America}

\emailAdd{j.panduro.vazquez@cern.ch}

\abstract{The ReadOut System (ROS) is a central part of the data acquisition (DAQ) system of the ATLAS Experiment at the CERN Large Hadron Collider (LHC). The system is responsible for receiving and buffering event data from all detector subsystems and serving these to the High Level Trigger (HLT) system via a 10~GbE network, discarding or transporting data onward once the trigger has completed its selection process. The ATLAS ROS was completely replaced during the 2013-2014 experimental shutdown in order to meet the demanding conditions expected during LHC Run~2 and Run~3 (2015-2025). 
The ROS consists of roughly one hundred Linux-based 2U-high rack-mounted servers equipped with PCIe I/O cards and 10 GbE interfaces.
This paper documents the system requirements for LHC Runs~2 and 3 and the design choices taken to meet them. The results of performance measurements and the re-use of ROS technology for the development of data sources, test platforms for other systems, and another ATLAS DAQ system component, namely the Region of Interest Builder (RoIB), are also discussed. Finally performance results for Run~2 operations are presented before looking at the upgrade for Run~3.}

\keywords{Data acquisition concepts, Data acquisition circuits}

\arxivnumber{insert} 

\begin{document}
\maketitle
 \flushbottom
 
\pagenumbering{roman} 
\setcounter{page}{2}

\clearpage
\section{Introduction}
\label{sec:intro}

The ATLAS~\cite{Aad:2008zzm} general purpose particle detector at the CERN Large Hadron Collider (LHC) is used to study a wide range of physics processes produced by proton or ion collisions. To maximise discovery sensitivity ATLAS employs a trigger system featuring both fast electronics and advanced software algorithms to select interesting events for further analysis. In order to both search for rare physics processes and better understand and measure more well known phenomena ATLAS must record and analyse a large number of particle collisions. To this end the LHC brings its beams into collision once every 25~ns, at which time multiple interactions between crossing beam particles may occur. Data generated by the resulting particles travelling through the ATLAS detector are then collected and analysed by the ATLAS trigger and data acquisition (TDAQ) system~\cite{Aad:2013tqj}~\cite{ATLASTDAQ:2016pov}~\cite{Panduro_Vazquez_2017}. 

While the TDAQ system performed well in LHC Run~1 (2010-2012)~\cite{ATLASTDAQ:2016pov}, upgrades to both the LHC~\cite{lpc} and the ATLAS detector itself during the 2013-2014 experimental shutdown made it necessary to redesign some parts of the infrastructure to handle even higher rates of data processing. The TDAQ upgrade also made it possible to satisfy new requirements driven by Run~1 operational experience as well as extending the performance reach of the system to both meet the needs of Run~2 and satisfy the expected requirements for Run~3. 

In this paper the upgrade of the Readout System, or ROS, during the 2013-2014 shutdown is discussed. The system itself operated successfully during LHC Run~2 (2015-2018) and remains the backbone of ATLAS readout in LHC Run~3 (2022-2025) alongside the newly introduced FELIX system~\cite{PanduroVazquez:2784274}. The design of the original Run~1 ROS is presented in detail in other publications~\cite{ATLASTDAQ:2016pov, DFReadout:ROBIN, Kugel:2009via, Crone2010534}. The Run~2 and 3 ROS consists of roughly one hundred Linux-based 2U-high rack-mounted servers, each equipped with either one or two PCIe I/O cards and two dual-port 10 GbE network interface cards (NICs).  The FPGA-based PCIe I/O cards~\cite{Engel:2683612}~\cite{borga}, developed by the ALICE Collaboration, are programmed with ATLAS-specific firmware, implementing the 'RobinNP' design. The system provides connectivity to approximately 2000 point-to-point optical links, often referred to as S-LINKs or S-LINK channels (as the S-LINK~\cite{BijSlink1997} protocol is used), conveying ATLAS event data. The RobinNP  buffers data from events accepted by the first-level hardware-based trigger (L1). This facilitates data sampling by the software-based High Level Trigger (HLT), running more complex algorithms on a commodity server farm, and readout of event data not sampled for HLT processing upon HLT accepts. The ROS is therefore the primary interface and buffer between the fast electronics of the detector readout and the commodity-hardware-based system downstream. 

The ROS was redesigned based on lessons learned during LHC Run~1, while also taking advantage of recent advances in technology. The redesign has made it possible both to meet the challenges of LHC Run~2~\cite{lpc} and, with the server update discussed in Section~\ref{sec:run3}, the expected requirements for Run~3. The following sections cover the motivation for upgrading the ROS in more detail, before moving on to look at the design itself. Performance results are presented, both from the laboratory and in Run~2 data taking, before a brief overview of developments for Run~3.

The design of the new ROS came against a backdrop of a move towards more common solutions and technology re-use in the field of DAQ. In this spirit, a number of different applications of ROS technology were investigated during Run~2 and beyond. In this paper, three specific development cases will be discussed: the software-based Region-of-Interest Builder (RoIB)~\cite{Abbott:2016} for Run~2 and the QuestNP and Dozolar data source facilities.

\pagenumbering{arabic} 
\setcounter{page}{1}

\clearpage
\section{The ATLAS TDAQ system}
\label{sec:daqsystem}

\begin{figure}[hbtp!]
	\centering
	\includegraphics[width=0.9\linewidth]{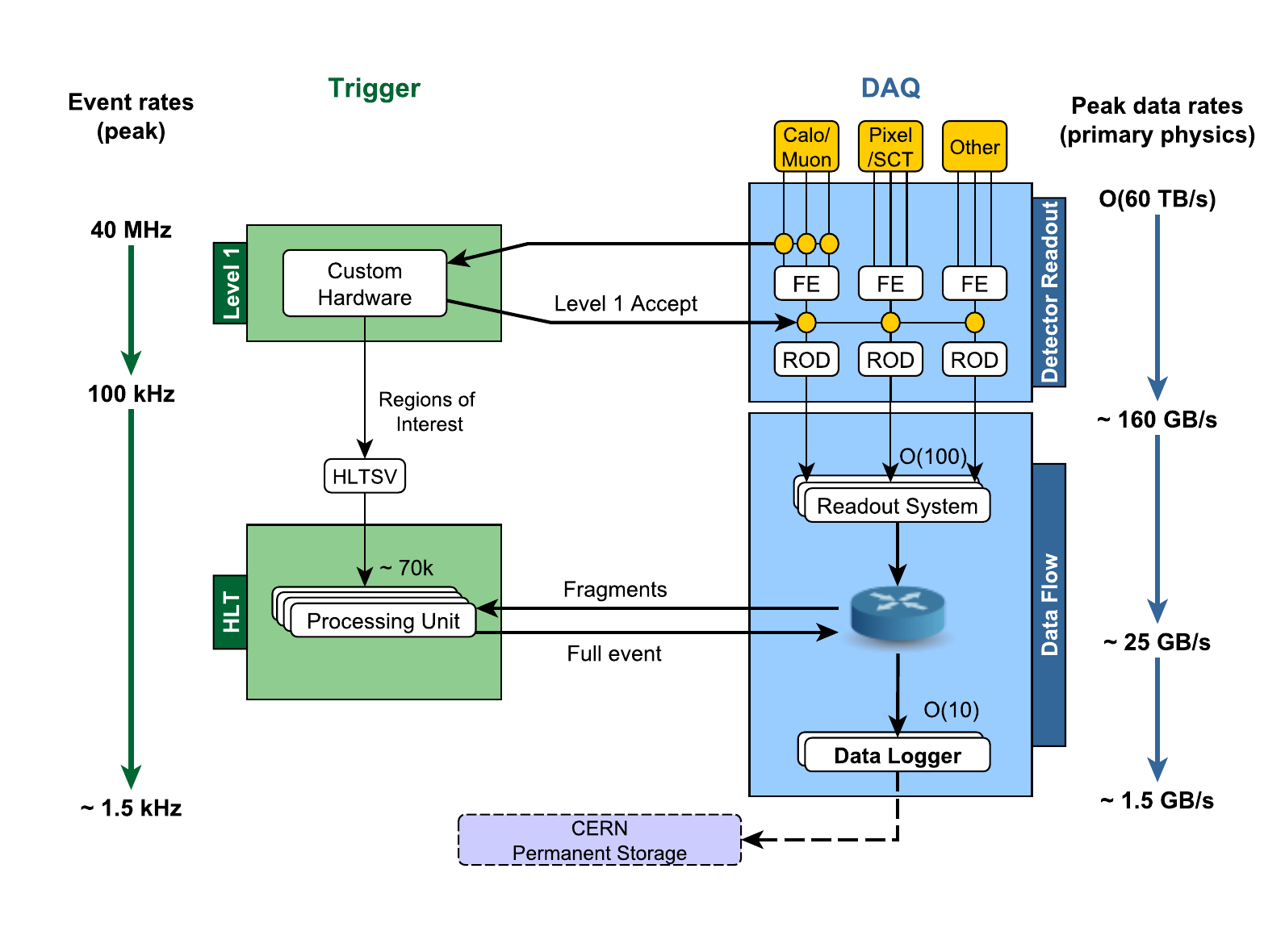}
	\caption{The ATLAS TDAQ system as implemented for LHC Run~2. The RoIB component was integrated into the HLT Supervisor (HLTSV) early in the run, so only the combined entity is shown. FE refers to front-end electronics and SCT refers to the Semiconductor Tracker detector. Requests from the HLT nodes to the ROS travel over the same network as the data sent in response. Note that the rates and throughput to permanent storage increased by approximately 50\% over the run period, so the values quoted indicate the initial expectations.
	}
	\label{fig:tdaqrun2}
\end{figure}

The ATLAS TDAQ system comprises both event selection and data transport and storage components. The system as implemented for Runs~2 and 3 can be seen in Figure~\ref{fig:tdaqrun2}. The first stage encountered by event data is the Level-1 trigger system (L1). This is implemented purely in hardware, using fast ASIC- and FPGA-based processing techniques. L1 decides on acceptance for further processing of each event on the basis of the electromagnetic and hadronic calorimeter as well as the muon system data. When an event is accepted (referred to later as an "L1 Accept" or L1A for short), information on regions with data satisfying certain criteria is collected and aggregated in the Region-of-Interest Builder (RoIB) before being passed on to the software-based HLT. This stage runs on a large computer farm based on commodity server hardware, and makes use of the Region-of-Interest (RoI) information to selectively sample the data from related regions in its first processing steps.

At the same time as the transfer of the Region-of-Interest information to the HLT, the
L1 system also signals the readout hardware for each detector subsystem
to process and transfer all information regarding the event in question to
the ATLAS Readout System (ROS). The subdetector front-end electronics transfer data to the
custom detector-specific hardware, known as readout drivers (ROD), for initial
processing before they are passed on to the ROS via optical fibres implementing
the S-LINK protocol. The ROS itself is a farm of servers housing custom-built 
electronics cards, called ROBIN in Run~1 and RobinNP in Runs~2 and 3, onto 
which are mounted memory modules for data storage and FPGAs to implement high 
performance data transfer logic. When data arrive in the ROS via S-LINK they 
are stored in the memory modules of the ROBIN/RobinNP. 

In LHC Run~1 the HLT system was implemented as two separate server farms, one
(known as the Level-2 or L2 trigger) dedicated to studying the data from interesting detector regions flagged by the L1 trigger system via the RoIB, and another (the Event Filter or EF) dedicated to studying complete events built by combining all data for events accepted after RoI analysis in the Level-2 trigger. This structure evolved into the combined system~\cite{zurNedden:2238679} described below in order to improve flexibility, scalability and reduce duplication of data requests to the ROS. More information on the Run~1 system can be found in a dedicated publication~\cite{ATLASTDAQ:2016pov}.

In LHC Runs~2 and 3, processing on the combined HLT farm is orchestrated by a single node, the HLT Supervisor (HLTSV), into which the RoIB functionality was integrated at the start of Run~2 (see Section~\ref{sec:pc_roib} for more details).
On receipt of RoI information, the HLTSV assigns the event in question to a specific farm node and monitors the status of processing, re-assigning the task to other nodes in case of problems (in which case the event is automatically accepted, but is flagged as requiring further debugging). On each HLT node, a process called the Data Collection Manager (DCM) sends requests to selected ROS servers to sample some or all of the data stored in their buffers on the basis of the RoI information received from the HLTSV. The ROS then serves this data over an Ethernet network with a maximum aggregate throughput of 40 Gb/s (via four 10 GbE links). Should the HLT decide to accept an event for permanent storage, the DCM will request all remaining data for that event from the ROS servers to transfer to permanent storage. At this point, the HLTSV sends a signal to all ROS servers to delete the relevant data from their buffers. Typically, these signals are grouped such that one hundred are sent at a time to optimise network use and processing efficiency. Should an event not pass HLT selection, the HLTSV will similarly request deletion from the ROS as soon as the decision is made (to be included in the grouping mentioned above). The final 'Data Logger' stage of the chain, which is responsible for packing accepted events into larger files and transferring them to permanent storage, is more commonly called the SubFarm Output (SFO).

\clearpage
\section{Motivations for upgrade}
\label{sec:motivation}

\subsection{LHC Plans through to 2022 and ATLAS requirements evolution}
\label{subsec:requirements}

Upgrades of the LHC during the period from 2013 to 2022 have allowed and will allow it to deliver collisions with both higher centre-of-mass energies and intensities. During the 2013-2014 experimental shutdown (i.e. at the end of Run~1) the centre-of-mass energy of proton-proton beam collisions increased from 8~TeV to 13~TeV. During Run~2, the peak collision luminosity increased to beyond double the original design value of $1\times10^{34}\mbox{ cm}^{-2}\mbox{s}^{-1}$. This peak luminosity is expected to be maintained for Run~3, which started in 2022, with an increase in the collision centre-of-mass energy to 13.6~TeV.

In addition to the evolving collision environment, changes to the ATLAS experiment itself have also placed new requirements on the readout system. In 2013-2014, changes to the HLT architecture resulted in the total HLT processing time for event data now determining the time during which data from an event has to be retained in the ROS~\cite{Panduro_Vazquez_2017}. This is considerably longer than in Run~1, where the determining factor was only the latency of the first stage of HLT processing and the building for events accepted by that stage. The longer HLT processing times and expected larger event sizes in Run~2 drove the need for larger buffering capacity in the ROS. Request loads were themselves also expected to increase as HLT algorithms and the events under study grew in complexity. The projected trigger request rates per ROS in Run~2 varied on scenario, but the maximum expected was 25~kHz per readout link and 50~kHz per ROS PC. In practice the rates significantly exceeded these predictions (see Section~\ref{subsec:run2_perf}). These increased rates are expected to also be required in Run~3.



Along with the changes to the HLT, the rate at which the L1 trigger accepted events would also increase from 75 kHz in Run~1 to nearer the ultimate design goal of 100 kHz~\cite{ATLASTDAQ:2016pov} in Run~2. On the detector side, a number of new subdetector systems were also integrated for Run~2, while others expanded the number of output channels required to be read out. In total the number of links was estimated to be increasing from \textasciitilde 1600 in Run~1 to \textasciitilde 2000 in Run~2. With the exception of the readout of the replaced Muon Small Wheels~\cite{Kawamoto:1552862}~\cite{PanduroVazquez:2784274}, this operating environment has been maintained in Run~3, with any significant additional readout paths being serviced by the new FELIX system.

All of the factors above combined meant that the ROS would have to service a larger number of readout links and sustain larger input and output rates from each part of the system in Runs 2 and 3 compared to Run~1. Additionally, at a system level, the size of the memory buffer per link in the ROS constrains the average HLT processing time and, as a consequence, the HLT farm size in terms of parallel processes. Because of this, a substantial increase in the buffering capabilities was deemed essential.

At first, consideration was given to simply expanding the existing system. However, this posed a number of problems. There were insufficient numbers of the original ROBIN cards to provide the additional channels and spares. It was not viable to assemble more of these cards due to the design being some 10 years old at the time. Many of the original components were obsolete and no longer readily available on the market. Also, as mentioned, the sizes of the ROBIN memory buffers were not expected be large enough for Run~2 operations due to the new HLT architecture, thus limiting the overall size of the HLT. Ongoing compatibility of the ROBIN cards, which have a 64-bit PCI interface, with contemporary servers was also an unavoidable issue. Furthermore, the aforementioned increase of the maximum L1 trigger accept rate was expected to load the on-board PowerPC CPU of the ROBIN, which has to process all the bookkeeping associated with the received event data and with all data and delete requests,  more heavily. Finally, the rack space available for expanding the system was limited such that it would prove difficult to expand by the required amount to guarantee meeting the performance targets. Given the size and bus requirements of the ROBIN cards a simpler solution, increasing the density by using servers smaller than the 4U high models used at the time, was also not considered viable.

Taking all of these factors into account, studies with realistic rates and conditions suggested that the performance would indeed be insufficient to meet the Run~2 request rate requirements, even when using 10 GbE network interfaces instead of the 1 GbE interfaces deployed in Run~1. 
To ensure that the ROS would provide the performance required for Runs~2 and~3 it was thus concluded that the system should be entirely replaced with a new implementation. As an added requirement, it was specified that the system should also be more compact than the previous generation, so as to allow more capacity for future expansion if needed for Run~3. Meeting this requirement has meant sufficient rack capacity has been retained for the installation of the FELIX platform alongside the ROS for Run~3 without major rack re-organisation.

\subsection{Design Overview of the Upgraded ROS}
\label{subsec:design_overview}

The upgraded ROS system was to be identical in role to its predecessor and share the majority of its functional components. The design of the system sought to make use of up to date technology to eliminate many of the known bottlenecks in the old system. Principal among these changes was the overhaul of the management mechanism for controlling transfers across the PCI bus (now PCIe in the new system) and exploiting the substantial performance gains made in commodity CPU performance, thus making it possible to move all of the operations previously performed in the ROBIN PowerPC into the host CPU. As a result of the redesign of the internal communication mechanisms used by the software, featuring increased use of interrupts and inter-thread signalling, it was possible to minimise costly polling for all interactions with the hardware.

In the following sections the key functional and design elements of the ROS will be described and discussed in more detail.
%
%


\clearpage
\section{Hardware}
\label{sec:hardware}

\subsection{RobinNP}

The RobinNP is based on a custom I/O board designed by the ALICE Collaboration, known as the Common Readout Receiver Card, or C-RORC~\cite{Engel:2683612}. The card was designed by ALICE for use both in their readout system and high level trigger, and came to the attention of ATLAS as part of the technology survey for potential host hardware for the RobinNP. Given the clear synergies and aligned timescales, the decision was made to adopt the C-RORC as the platform for the RobinNP and manage the production as a joint effort between the two experiments. The ATLAS-specific functionality (RobinNP) was then implemented in firmware on the C-RORC hardware platform.

\begin{figure}[hbtp!]
	\centering
	\includegraphics[width=\linewidth]{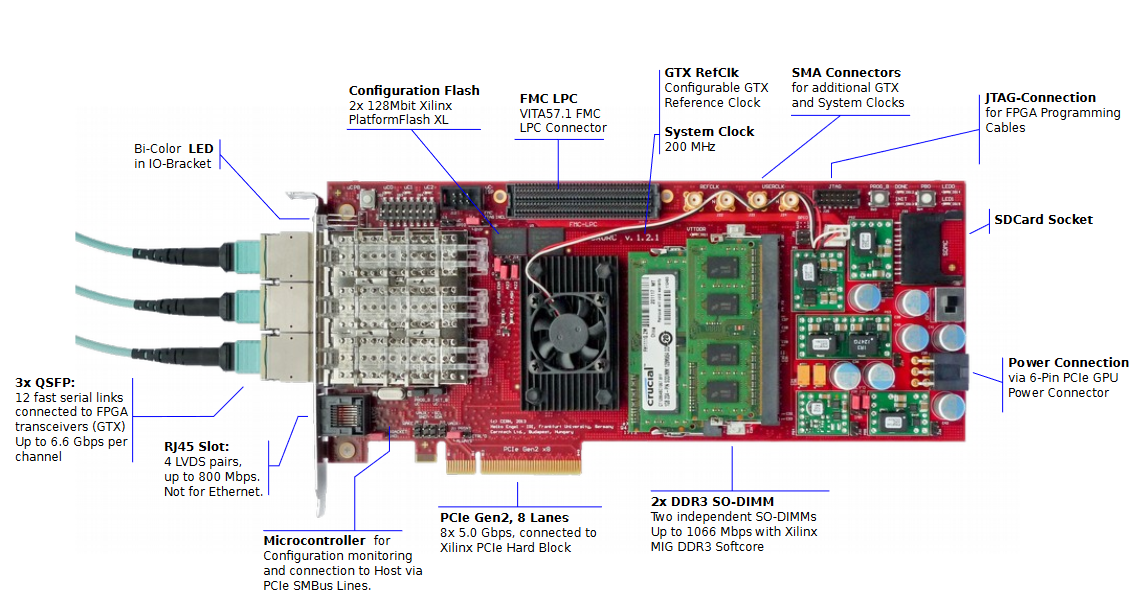}
	\caption{The ALICE C-RORC, used by ATLAS as the platform for the RobinNP. The Xilinx Virtex-6 FPGA is located centrally underneath the large black heat sink and fan.}
	\label{fig:robinnp}
\end{figure}

The C-RORC (which can be seen in Figure~\ref{fig:robinnp}) is based on PCIe technology (Gen 2 with up to 8-lanes), for a maximum theoretical data transfer rate of 4 GB/s. On board is a Xilinx Virtex-6 FPGA (XC6VLX240T-2FFG1156C)~\cite{virtex6} and two SO-DIMM slots capable of hosting up to 8 GB of DDR3 RAM each (for the RobinNP implementation each slot is populated with 4 GB). The C-RORC has three Quad-SFP (QSFP) cages, each capable of supporting four S-LINK connections via MPO connectors for a total of 12 links per card. The C-RORC has no on-board PowerPC (or similar) for process management, in keeping with the design paradigm for the RobinNP (as described in Section~\ref{subsec:design_overview}).

\subsection{Host Server in Run~2}

When selecting a host server for the RobinNP a number of factors were considered. A machine was needed which could host at least 2 C-RORC cards with added space for 4$\times$10 GbE connectivity. The throughput provided by two of these links is needed for data taking, while the other two provide redundancy. The machine itself was to be as compact as possible to maximise density, preferably with lower overall power consumption than the Run~1 equivalent. Finally, consideration was given to the layout of the PCIe and memory data paths on the motherboard to ensure that all peripheral hardware had equal priority in terms of CPU access. This factor was particularly significant when evaluating dual CPU motherboards, which were ruled out after benchmark tests with several candidates. In normal operation the servers were to be operated in network-boot mode, but a local disc was still included for incidental storage and use in test laboratory installations.

With all requirements considered, the final specification for the Run~2 ROS server was as follows:

\begin{itemize}
	\item Supermicro X9SRW-F motherboard (single CPU socket)~\cite{ROS:SupermicroX9SRW-F} with riser card with 4 PCIe 8 lane slots (RSC-R2UW-4E8) 
	\item Intel(R) Xeon(R) CPU E5-1650 v2, 3.50~GHz (6 physical cores with hyperthreading)~\cite{Intel:E5-1630V2}
	\item 16 GB DDR3 RAM (four 4 GB DDR3 modules, 1866 MHz, ECC) (DIMM DDR3-1866 4 GB ECC reg Micron MT18JSF51272PDZ-1G9K1)~\cite{ROS:RAM}
	\item 2x Mellanox ConnectX-3 dual port 10 GbE Network Interface Card (NIC)~\cite{ROS:MellanoxNIC}
	\item 2U server chassis, Supermicro SC825TQ-R500WB~\cite{ROS:SupermicroSC825TQ-R500WB} with redundant power supply
	\item 64 GB SSD for incidental storage (Sandisk X110 SSD 64 GB SATA)~\cite{ROS:SSD}
\end{itemize}

\clearpage
\section{Firmware design}
\label{sec:firmware}

The RobinNP firmware is similar in function to that of the ROBIN, but makes use of modern FPGA technology to implement more logical blocks into firmware rather than dedicated integrated circuits, while also achieving much higher channel density. In this section an overview of the design will be presented, followed by a discussion of the key elements.

\subsection{Overview, Organisation and Interfaces}

The Virtex 6 FPGA on-board the C-RORC is directly connected to the input optical QSFP modules, the DDR3 memory banks and the PCI express interface. All the interaction logic between components is implemented in the FPGA firmware.

\begin{figure}[hbtp!]
	\centering
	\includegraphics[width=0.9\linewidth]{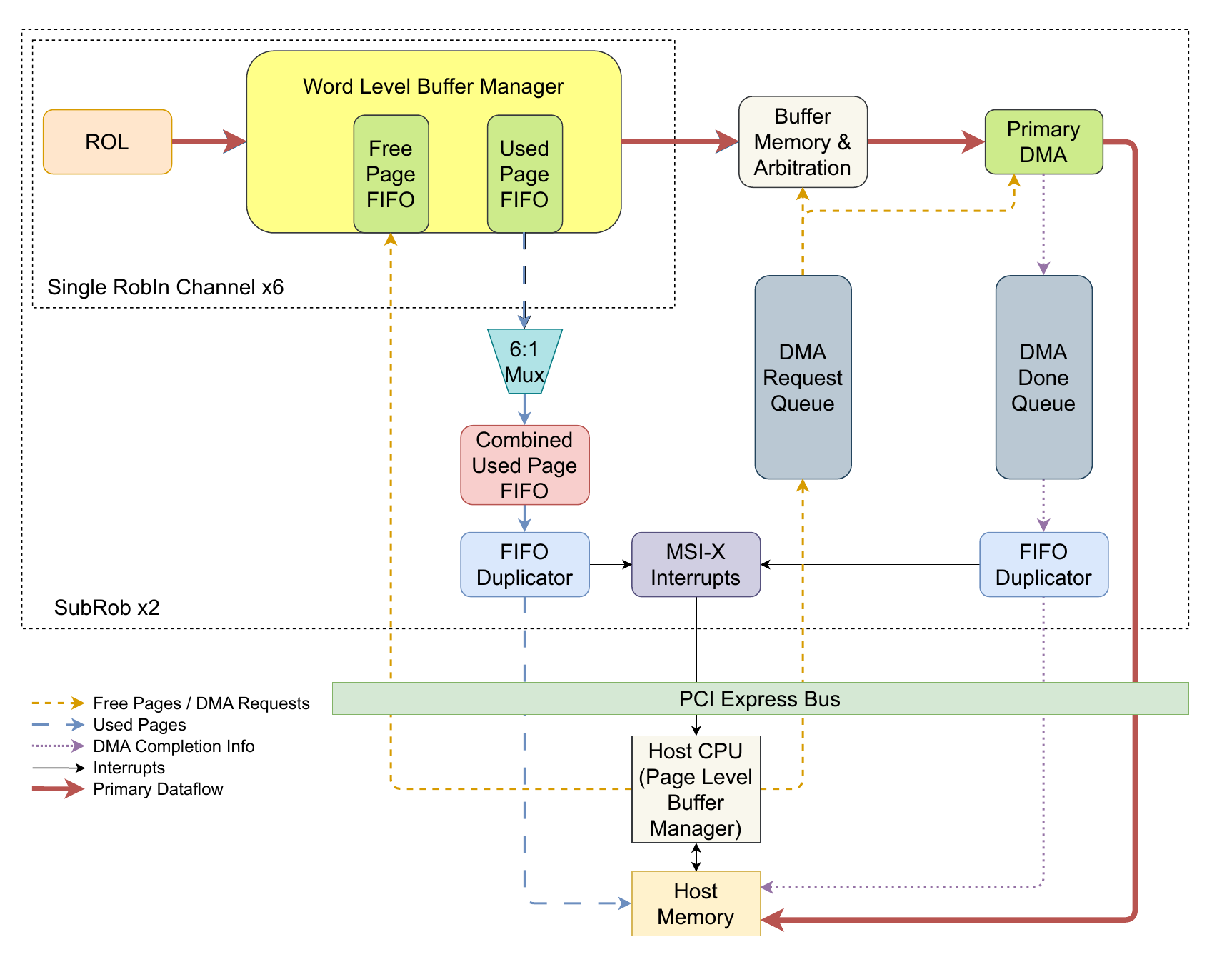}
	\caption{A block diagram of the complete RobinNP firmware, showing a SubRob (two copies) servicing six input links through common logic before transfer of data across the PCIe interface via DMA.}
	\label{fig:areducedblockdiagram}
\end{figure}

Given that the C-RORC supports two memory banks, it was decided to split the firmware into two identical sets of functional modules, known as `SubRobs'. A block diagram of the overall design is shown in Figure~\ref{fig:areducedblockdiagram}. Each SubRob supports six input S-LINK channels, shown as Readout Links or 'ROL', the data from which pass through common memory access arbitration and request management logic. Each SubRob has its own DMA engine and external endpoint in host server memory.

\subsubsection{S-LINK}
\label{subsubsec:s_link}

S-LINK~\cite{BijSlink1997} was developed at CERN to provide a standard stream protocol for detector data readout links. It was designed to look much like a FIFO, with data inserted at one end and taken from the other. From the user perspective S-LINK exposes a 32 bit interface with a write enable bit and flow control. This interface was initially developed with dedicated hardware, but in more modern implementations such as the RobinNP it exists as a logical interface within the FPGA firmware. Various physical layers have been used over time, but the 2~Gb/s fibre version called HOLA predominates and is now itself often referred to as S-LINK. The core logic of S-LINK interfaces connects to the physical fibre layer via a Texas Instruments Transceiver TLK2501~\cite{tlk2501}. The TLK2501 handles 8B/10B encoding, Serialisation/De-Serialisation (SerDes) and data synchronisation of the physical link. With recent evolution of FPGA I/O it became possible to support the speed and functionality of the TLK2501 within the FPGA itself without dedicated hardware. A 'logical' TLK2501 was therefore written with all the functionality of the TLK2501, with similar interfaces, so it could be connected to the S-LINK core and optical transceivers without changing their interfaces. This functional block is referred to in this document as the TLK. 

In the case of the RobinNP, the three on-board QSFPs each have 4 duplex links, providing a total of 12 input channels per card. The data bandwidth of the S-LINK is determined by the S-LINK clock. The original ROBIN was clocked at 40 MHz giving a bandwidth of 160 MB/s but increasing the clock to 50 MHz would allow up to 200 MB/s. The RobinNP S-LINK clock operates at 64 MHz, so any detector wishing to make use of the higher bandwidth can do so if their electronics permits. Furthermore, by increasing the fibre reference clock from 100 MHz to 125 MHz it is possible to further increase the bandwidth to 250 MB/s. This is also supported by the RobinNP should it be required by any given detector. In practice many detectors are limited to the 160 MB/s case by legacy electronics, leading to some upgrades being made as presented in Section~\ref{subsubsec:limitations}.

\subsubsection{PCIe Interface}

To minimise development overhead, a commercial PCIe~\cite{pcie} core~\cite{plda} (including DMA management engines), was used to manage all interface operations. Numerous control and status registers were mapped into the host system's memory space across the interface to facilitate management of all hardware components. While the C-RORC hardware supports up to PCIe Gen 2x8, the dataflow requirements of the ROS did not indicate that full throughput would be required. Therefore, to make it simpler to fit all necessary logic into the FPGA while achieving timing closure, it was decided to use a lower throughput interface in the design. In practice it is possible to build the firmware to support either PCIe Gen 1x8 (which is used in the production ROS) or PCIe Gen 2x4 (to maximise the number of compatible motherboard slots). Both builds deliver the same throughput, which has in practice been measured at 1.6 GB/s, accounting for all overheads. A full Gen 2x8 build could be considered in future if extra performance is required.

\subsection{Event Fragment Formatting and Quality Management}

By convention, all incoming fragment data should contain pre-defined control words at the beginning and end. The first part of each event fragment contains a header~\cite{Bee:683741} with various pieces of information, including the L1 trigger ID (L1ID) which acts as the overall event identifier. The RobinNP firmware performs data quality checks on all incoming fragments before passing them on to the downstream stages of the system. The first check is on the presence of the pre-defined control words corresponding to the beginning-of-fragment and end-of-fragment markers. The fragment also has to be long enough to contain the L1ID, as without it the fragment cannot be indexed correctly. Errors are marked in the stream for processing later by a component known as the `Buffer Manager'. Additionally a CRC word~\cite{crc} is appended to the data, allowing for later checks to ensure the data are not corrupted during transit through the RobinNP. The RobinNP software can additionally be configured to dump copies of any fragment flagged with such errors to disc for aid debugging either of the ROS itself, or the detector electronics.

\subsection{On-board Memory Management}

\begin{figure}[hbtp!]
	\centering
	\includegraphics[width=\linewidth]{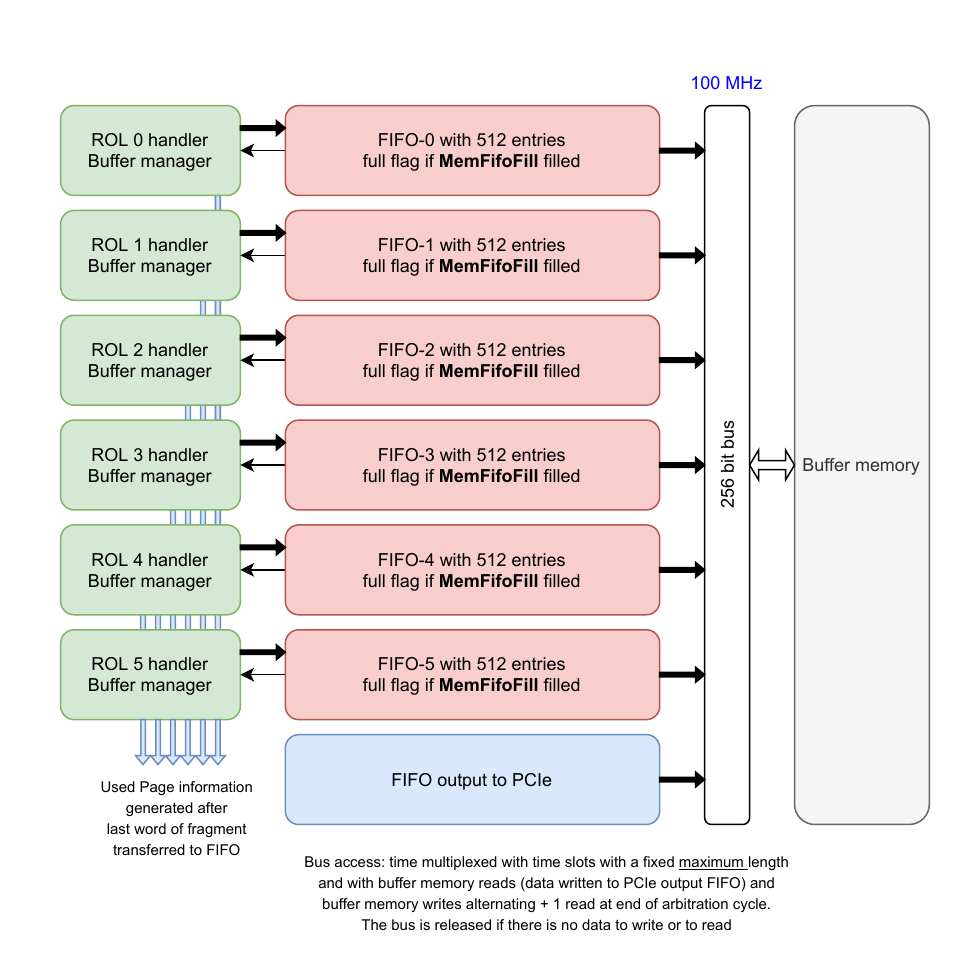}
	\caption{RobinNP memory access and arbitration schematic. If the number of items stored is equal to or larger than a programmable level (MemFifoFill), this results in the full flag being raised. The test Input Channel is not shown.}
	\label{fig:arbiter}
\end{figure}

Each RobinNP input channel features a Buffer Manager module as an interface to one of the two on-board memory modules. To ensure an even distribution of data, each memory module has an input from the Buffer Managers of six input channels. Each memory module is logically subdivided into equal parts among the channels, for a total of 0.67 GB storage per channel. Access to the memory is multiplexed among the six Input Channels (Memory Write), one Output Channel (Memory read) and one test Input Channel (Memory Write), as shown in Figure~\ref{fig:arbiter}. The low level control of the memory is implemented using a Memory Interface Generator (MIG) core provided by Xilinx~\cite{mig}. In the current implementation each memory module is able to maintain a read bandwidth of the order of 900 MB/s, which given there are two modules is sufficient to saturate the Gen 1x8 PCIe interface.

\subsubsection{Storage of Incoming Data}

As data arrive on each input channel, the relevant Buffer Manager writes the data into memory. The Buffer Manager has two interfaces to the host system which control where data are stored. The memory is divided into discrete units or `pages' of configurable size (typically 4~KiB). Each page is enumerated, providing a token for the host system to use to refer to a given memory location. When the memory is empty, the host system holds all unused page tokens in a stack structure in software. To make memory locations available for data storage, the software removes page tokens from its unused store and writes an entry to a firmware FIFO known as the `Free Page FIFO` (FPF) to indicate that a given location is available. One such FIFO exists per input data channel. 

Data which come into the Buffer Manager are then written to the next free page indicated by the FPF, from which the corresponding entry is then removed. When a memory page is full, or the end of an incoming event fragment is reached, the address of the last word written is copied to the `Used Page FIFO' (UPF) for the relevant channel, for which the contents are then transferred to host memory via a mechanism known as the FIFO duplicator (see below for a full description). The software is then able to store the tokens associated with these used pages in a separate structure. The FPF and UPF thus allow the software on the host system to manage the allocation of memory for incoming data.

\subsubsection{FIFO Duplicator}

\begin{figure}[hbtp!]
	\centering
	\includegraphics[width=0.9\linewidth]{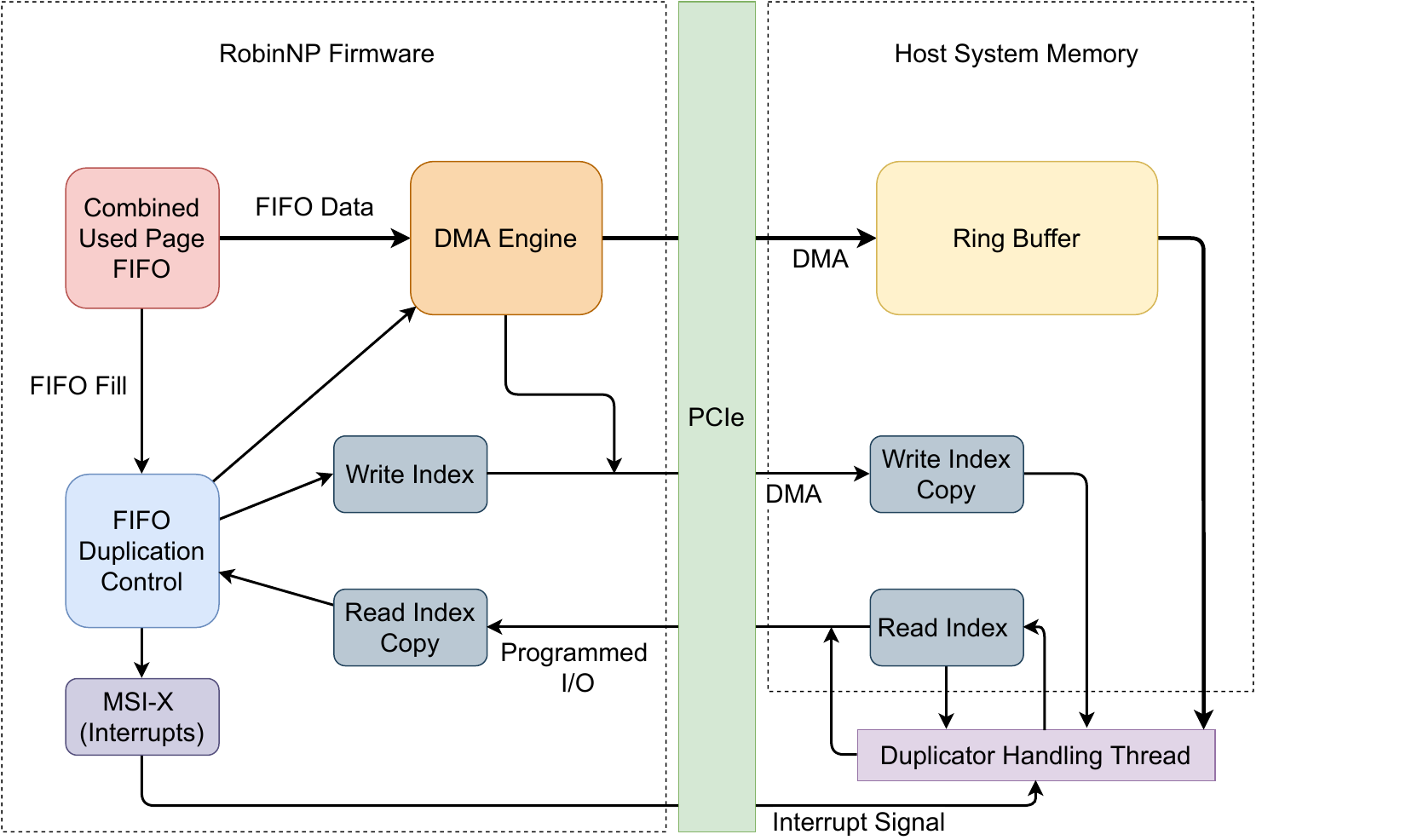}
	\caption{Schematic of the FIFO Duplicator Protocol.}
	\label{fig:fifoduplicator}
\end{figure}

FIFO duplicators are a mechanism that makes it possible to maintain instances of FIFOs in host memory which can be accessed by both firmware and software. The aim is to avoid the software having to perform costly read operations across the PCIe interface, which would be required if the FIFOs were to be implemented purely in firmware. The duplicator works by taking entries from the firmware version of the FIFO and moving them into a ring buffer in host memory, managed by software. As data flows through the system the ring buffer is kept populated by the firmware, with automatic DMA transfers from the firmware FIFO. The firmware and software share management of the read and write pointers of the ring buffers such that the firmware cannot overwrite unread data and the software cannot read beyond the populated area. A schematic of the logical blocks underpinning the duplicator mechanism is presented in Figure~\ref{fig:fifoduplicator}.

With duplicators, PCIe read operations are replaced by much less costly PCIe write operations, making it possible for the host software to manage both the indexing and DMA functions previously housed within the PowerPC of the Run~1 ROBIN. The Duplicator mechanism was further improved by notifying the host of new data using interrupt driven notifications, thus avoiding the use of tight polling loops to check completion of transfers in situations where this was likely to waste CPU cycles. Message Signalled Interrupts (in particular MSI-X) were used, which unlike traditional interrupts send a message to the host CPU rather than asserting one of the PCI interrupt lines. The result is the same except there are many more MSI-X interrupts available (2048) than the four on a standard PCI system, thus providing for a much more flexible and targetted design.

\subsubsection{Readout of Outgoing Data}

\begin{figure}[hbtp!]
	\centering
	\includegraphics[width=\linewidth]{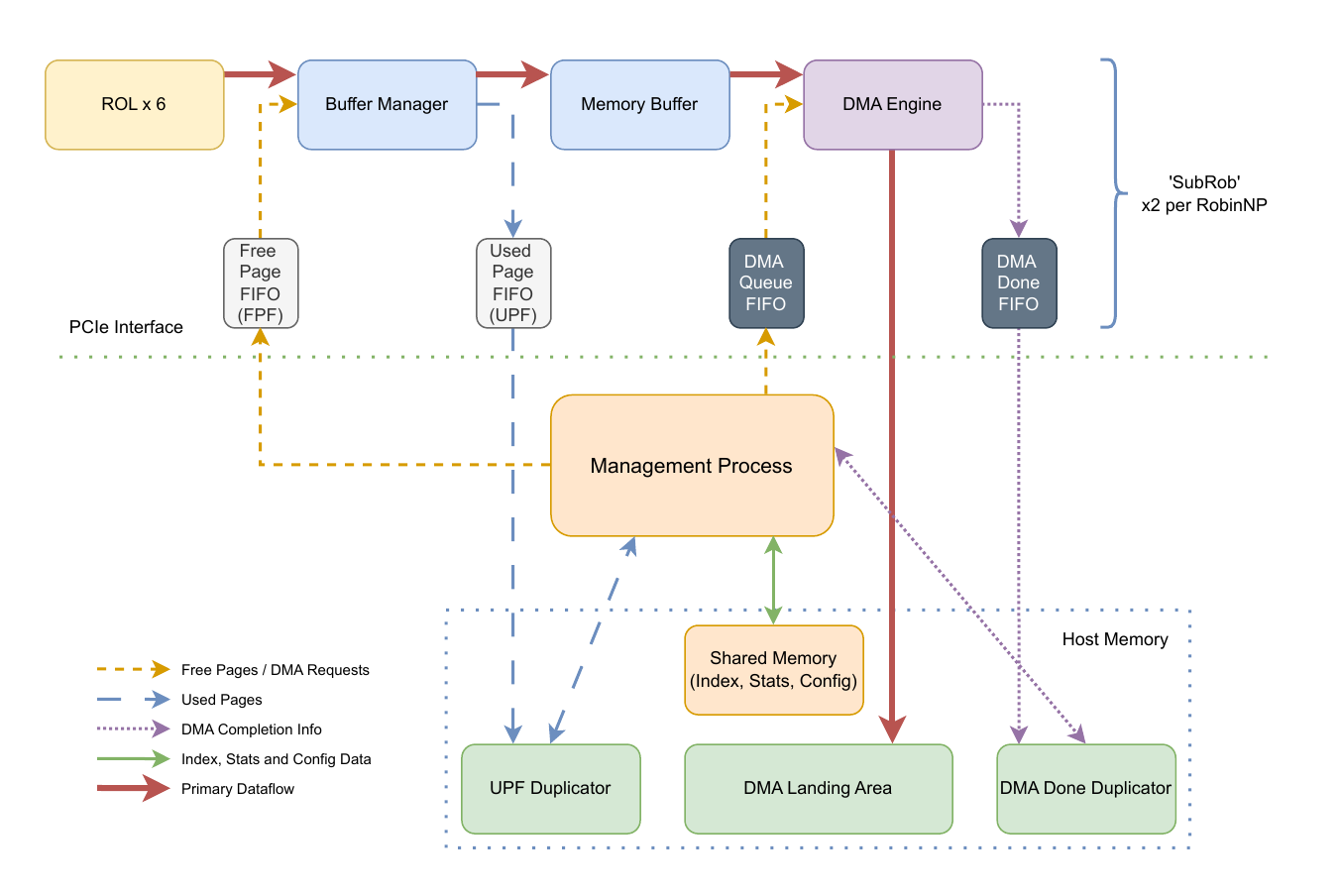}
	\caption{Run~2 ROS firmware/software interaction.}
	\label{fig:run2datatransfer}
\end{figure}

The transfer of data from the RobinNP memory buffers to host server memory is done using a DMA transfer. A basic schematic for the read (and write) mechanism is shown in Figure~\ref{fig:run2datatransfer}. The DMA transfer is initiated by the software writing a DMA descriptor to a dedicated firmware FIFO (DMA Queue FIFO). This initiates a read from the memory and also triggers a DMA transfer of the resulting data to the host memory. When the DMA transfer has been completed an entry is written to a `DMA done' FIFO in host memory, managed by the same kind of duplicator mechanism as used for the `Used Page FIFO'. If armed, an interrupt is asserted when the ring buffer of the 'DMA done' FIFO duplicator is in a non-empty state. Thus, the software is alerted to the fact that a data payload from memory is available for processing. More information on the software portion of these features will be presented in Section~\ref{sec:software}.

\subsection{Hardware Management and Health Monitoring}
\label{subsec:mgmt_mon}
The C-RORC card hosts a microcontroller for the management of on-board components. Access to the management bus is via two I2C interfaces~\cite{i2c}, one connected to the FPGA and another directly to the host motherboard via SMBus~\cite{smbus}. Through the first interface it is possible to read device information for, among other things, the FPGA, memory modules and QSFPs, including operating temperatures and voltages. It is also possible to perform more complex operations such as managing the on-board crystal oscillator. The SMBus interface is used to trigger configuration of the FPGA from any one of the four FLASH sectors available. The FPGA can also be configured directly via JTAG~\cite{jtag} through a dedicated connector. The FLASH sectors can be pre-loaded with any required firmware images either over JTAG or via PCIe if the FPGA is temporarily configured with separate firmware solely for this purpose.

\subsection{Firmware Data Generator}
\label{subsec:datagen}
To facilitate the testing of RobinNP cards in scenarios where external S-LINK inputs are not available, the firmware includes a configurable data generator module, able to inject properly formatted fragments~\cite{Bee:683741} with pre-selected data patterns into each of the 12 on-board channels in parallel. Within each channel, the L1ID generated is set to zero when the generator starts to operate and increased monotonically with each generated fragment until the 32-bit field available for the ID wraps back to zero (configurable to occur at a smaller value) or the generator is restarted. Data can be generated at rates high enough to saturate the channel bandwidth (up to 250 MB/s each across 12 channels) for sufficiently large fragments. 

The size of the fragments to be generated can be chosen by the user, defining a data payload to which 14 words of header and trailer information are added. For example, a configuration which generated 100 word payloads would in fact generate 114 words of data. The pattern generated in the data payload is a counter, monotonically increasing for each 32-bit data word generated.

\clearpage
\section{Software design}
\label{sec:software}

The redesign of the ROS software infrastructure for Run~2 was influenced by a number
of new requirements. The opportunity was also taken to use modern techniques
to improve performance and code maintainability. The overall strategy was to replace
the old thread pool model with a new one based on simpler, single responsibility threads.
Such threads could then be more closely aligned with the firmware interface to better
support the interrupt-driven communication model. All inter-thread communication was
designed to be based on semaphores or signals, with the aim of eliminating any tight
polling loops. The software can be considered in terms of two layers: the base interface with the firmware, and the wrapper interfacing with the wider dataflow system. The higher level software integrates directly into the ATLAS data taking infrastructure, but lower level tools also exist to test the system in stand-alone mode. In this section, both high and low level layers will be described in detail, including key components of the software suite, such as the device driver.

\subsection{Base Interface with Firmware}

The interface between the software and firmware consists of a series of firmware control registers and FIFOs with a return path via several instances of the FIFO Duplicator mechanism described in Section~\ref{sec:firmware}. Each firmware SubRob has its own set of dedicated software processing threads. Software processes can interact with the firmware by writing to the relevant control registers to trigger specific actions. The RobinNP memory buffers are subdivided into small blocks, known as pages, which are managed by the host software. The addresses of pages which contain no event data are provided to the firmware via a FIFO, managed by a dedicated thread. Once a page has been populated with event data by the board, the address is returned to the software via a FIFO Duplicator, also managed by a dedicated thread. Pages arriving over the duplicator are placed in an event index, implemented as a linked list. The L1ID of the event is used as the key for all indexing operations by way of a hash based on the lowest 20 bits.

\begin{figure}[hbtp!]
	\centering
	\includegraphics[width=0.8\linewidth]{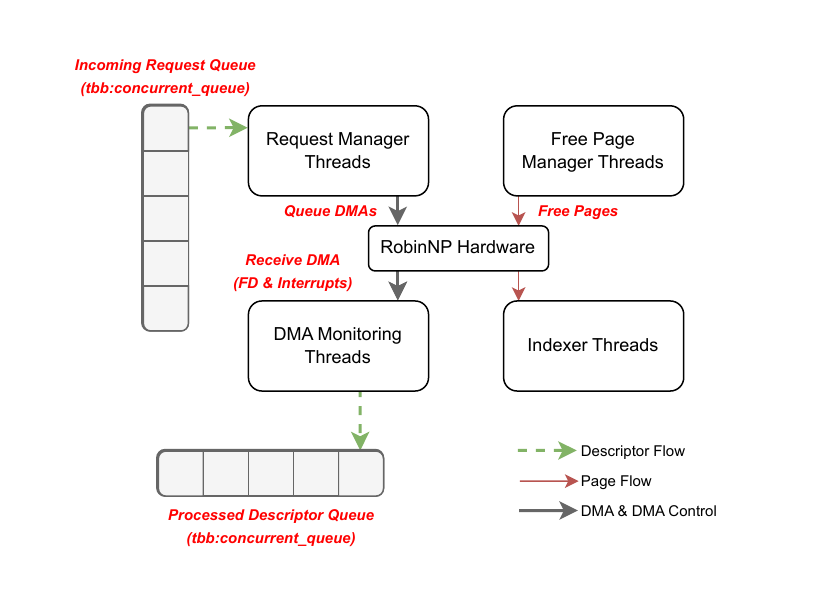}
	\caption{RobinNP core firmware interface threads. Each firmware SubRob has its own set of threads and FIFOs, thus for each RobinNP card there are two copies of the structure shown in the diagram.}
	\label{fig:corethreads}
\end{figure}

Event data requests arrive via a queue structure based on Intel threaded building blocks (TBB) technology~\cite{intel:tbb}. Incoming requests are processed by a dedicated thread, which triggers a DMA transfer from the RobinNP memory into a pre-allocated page in host memory. Completion of a DMA transfer is signalled to the software by the firmware by means of another FIFO Duplicator, curated by another thread. This final thread passes the location of the event data to the outgoing message system via another TBB-based queue. The core threads as described above are presented in Figure~\ref{fig:corethreads}.

\subsection{Interface with Downstream Dataflow}

The RobinNP core threads interface to the wider dataflow system by means of a separate thread layer which handles initial processing of data requests as well as collecting returned data for transmission downstream. The primary interface is managed by a pool of I/O threads based on boost::asio~\cite{boost:asio}. The threads can operate both to service incoming requests and transmit outgoing data, but no thread is designed to handle both request and transfer for the same event, thus differentiating them from the Run~1 implementation. Alongside a configurable number of the Async I/O threads, each ROS application instantiates a single Collector and Pending Handler thread. 

Throughout their passage through the system, requests are associated with `descriptors', which are unique tokens encapsulating all the information and data corresponding to a specific request. Each ROS application is initialised with a configurable number of descriptors that remains fixed during runtime. When the HLT requests data the Async I/O thread handling the query takes a free descriptor from the queue storing them, populates it with the request description, namely the channels to read out, and passes this on to the Pending Handler Thread. This thread checks whether the request can be satisfied and, if so, passes the descriptor on to the core threads for the SubRob in question. If a request cannot be immediately satisfied it is placed on a queue for later checking. A check will be triggered if either a specified timeout elapses, or an additional data or delete requests arrives. Should the data still not be available in the latter case and the timeout has not yet elapsed, the request will be re-queued. Should it not be possible to satisfy a request, a response is returned to the HLT indicating that the data are not available. 

\begin{figure}[hbtp!]
	\centering
	\includegraphics[width=0.8\linewidth]{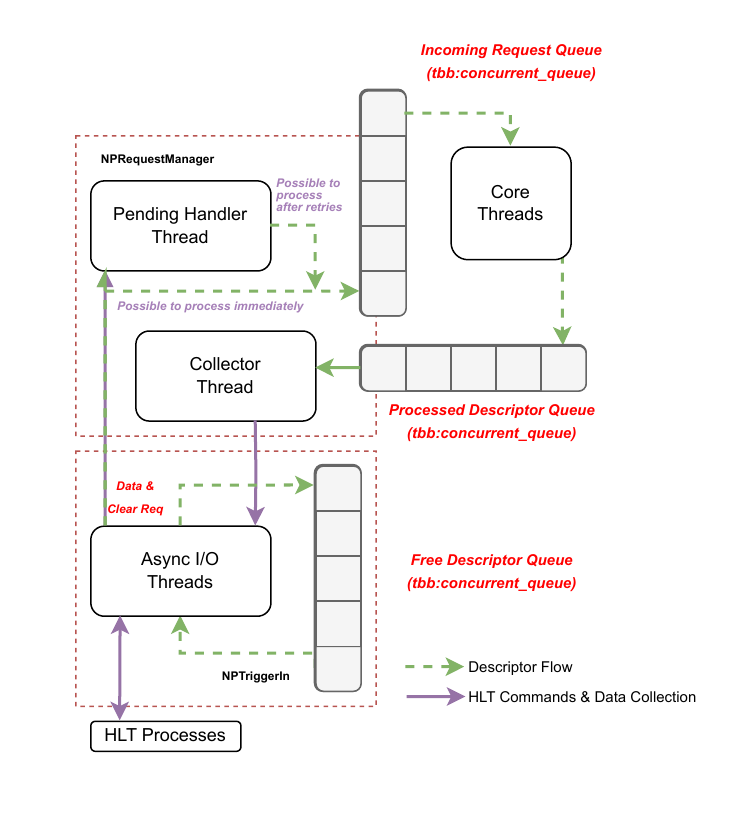}
	\caption{RobinNP request management threads, including Async I/O, Pending Handler and Collector threads. Also shown is the flow of commands and associated descriptors throughout the system. }
	\label{fig:rosthreads}
\end{figure}

For requests that can be satisfied, the core threads populate the descriptor with the location of the data read out from the RobinNP memory and pass it to an output queue (processed descriptors). For requests that can only be partially satisfied, i.e. some but not all requested links have data available, a response is sent to the HLT containing both the available data plus a notification of the links for which data could not be retrieved. This final HLT communication step is triggered by a collector thread, which takes completed descriptors from the output queue and checks that all readout for a given request is complete before passing the descriptor back to an Async I/O thread. From here, the data referenced by the descriptor are transferred to the HLT node from which the request came via the 10 GbE network links, before recycling the descriptor for re-use later. The thread structure is presented in Figure~\ref{fig:rosthreads}.

\subsection{Device Driver and Low Level Tools}

Interactions between all high-level software applications and the RobinNP firmware are mediated by a dedicated device driver. The driver is not responsible for low level control of the hardware beyond the initialisation, but simply allows the user software to map the base address register (BAR) into user space.

On startup, the driver allocates a pre-configured number of host memory pages for the purposes of DMA transfer, to be made available to high-level software on request. The driver also initialises the crystal oscillator on-board the RobinNP card to the correct frequency to drive the S-LINK connections. During normal operation, the driver manages the MSI-X interrupt feature of the firmware, exposing an interface though which high-level software may arm and mask interrupts, as well as signalling to any waiting threads when such an interrupt is received. In order to facilitate monitoring of the status of the hardware, the driver also produces an entry in the proc filesystem featuring dataflow information, link status reports and temperature and voltage values read over the I2C bus.

Beyond the driver, a suite of low level applications exists to facilitate testing and debugging of the RobinNP, from simple status monitoring to full dataflow emulation. 

\subsubsection{Operating System Compatibility}

The RobinNP software suite (and wider ATLAS DAQ infrastructure) were designed to work with 64 bit Linux Operating Systems. Due to ATLAS requirements, specific compatibility has been verified with Scientific Linux 6 and CentOS7, with work ongoing to add support for RHEL 9/AlmaLinux 9. Migration to newer operating systems will take place as needed, but no specific compatibility concerns exist.

\clearpage
\section{Performance}
\label{sec:performance}

In this section, an overview of the performance measurements performed with the upgraded ROS in the laboratory will be presented, followed by real operational performance data from Run~2.

\subsection{Laboratory measurements}

\subsubsection{Method}

Dedicated laboratory tests were performed to establish the maximum input event fragment rates and data request rates supported by a ROS server. The test setup consisted of a single ROS server connected via 10 GbE network connections to a number of servers running test software. Event fragments were generated by data generator modules built into the RobinNP firmware. The fragment size of the data generated is configurable under software control. However, the size cannot be changed during a run. The data generators run at the maximum possible rate dictated by either the available output bandwidth per generator (250~MB/s) or by backpressure due to the full flag of the FIFO receiving the generated data being raised. External data generators, consisting of 64-bit PCI cards with an FPGA, where each card can drive four optical links using the S-LINK protocol, have also been used in the past. Where the S-LINK bandwidth (see Section~\ref{subsubsec:s_link}) does not limit the rate similar results as with the internal data generators are obtained, with throttling of the rate achieved by means of the S-LINK XON-XOFF protocol~\cite{BijSlink1997}, driven by the full flag of the FIFO in which the data are received being raised.

The ROS server used in the test was equipped with two RobinNP cards, with all 12 internal data generators activated for each (i.e. 24 in total for the server). Data can be requested from the ROS server via the network from either all links at once or from a subset. Since in Run 2 and 3 requested data are cached within the HLT, the same data from a given link will not be requested more than once. This is different from the Run~1 system, where multiple requests for the same data were possible. 

\subsubsection{The ROSTester program}
\label{subsubsec:rostester}

The primary application used to drive the tests is known as ROSTester. This generates requests and receives event data. The measurements were done with four ROSTester instances running on two servers, each connected via two point-to-point 10 GbE links to the ROS server under test. The ROSTester program has an internal trigger generator that emulates L1 trigger accepts (generating L1IDs used in data and delete requests). ROSTester instances can issue either event data requests for only a subset of the channels, referred to as an L2 or RoI request (referring to data from a Region-of-Interest), or for all the data held by every Readout Link (ROL) not yet requested by an L2 request, referred to as an event building or EB request. It is also possible to send multiple L2 requests for a given event (without asking for the same data twice). These request patterns match the behaviour of the HLT in standard data taking.

For each emulated L1 trigger accept, the application decides whether to issue L2 and/or EB requests according to probabilities defined in its configuration. For an L2 request, the application chooses which ROLs to request at random (with each given equal probability) up to the pre-defined size of the L2 request, i.e. how many ROLs to include, also known as the RoI size. The number of outstanding requests is bound to a maximum (typically several hundred).  The rate at which L1IDs are generated is not allowed to exceed a configurable maximum. Once all requested data for an L1ID are received a delete request is generated to instruct the ROS to discard the associated data in its buffers. Delete requests are sent in blocks of 100 to the ROS server to optimise performance. The rate at which deletes occur is effectively analogous to the L1 Accept rate, which can be demonstrated to behave stably and realistically in tests (when managed by a well-chosen combination of ROSTester parameters) and with backpressure to the ROS causing RobinNP data generation to halt when needed. The  maximum possible delete rate can then be measured as a function of the L2 request fraction, the RoI size, the EB request fraction and the fragment size. 

Four ROSTester instances were used in parallel to reduce the CPU utilisation per ROSTester, avoiding that the measured delete rate be determined by ROSTester performance rather than ROS performance. Such an arrangement also better emulates the interaction with multiple HLT processing nodes that is expected in normal data taking. If one ROSTester program is used the L1ID increments by 1 for successive trigger accepts issued by the application. With four ROSTester programs running in parallel, the L1IDs generated in each application increment by 4 and start at a different value, such that requests for data associated with a given L1ID are only generated by a unique ROSTester application.


\subsubsection{Modelling ROS Utilisation}
\label{subsubsec:utilisation}

To model the overall load on a ROS in specific data taking conditions, hereinafter referred to as `Utilisation', numerous tests were performed with the ROSTester application and the results used to inform a mathematical description. To start, measurements of the delete (L1) rate (which as discussed above can be considered analogous to L1 Accept rate under these conditions) were performed with the version of the operating system and of the ROS software matching those used during data taking at the end of Run 2. The RobinNP's internal data generators were used to produce the fragment streams necessary. The fragment sizes used were 300, 350, 400 or 450 4-byte words. For each fragment size the L2 and EB request fractions were varied as part of the test alongside the RoI size. For EB requests data were requested for all ROLs for which data were not already requested by an  L2 request, whereas for L2 requests either 2, 3 or 4 ROLs were requested (i.e. the RoI size).


\begin{figure}[h]
	\centering
	\includegraphics[width=1.0\linewidth]{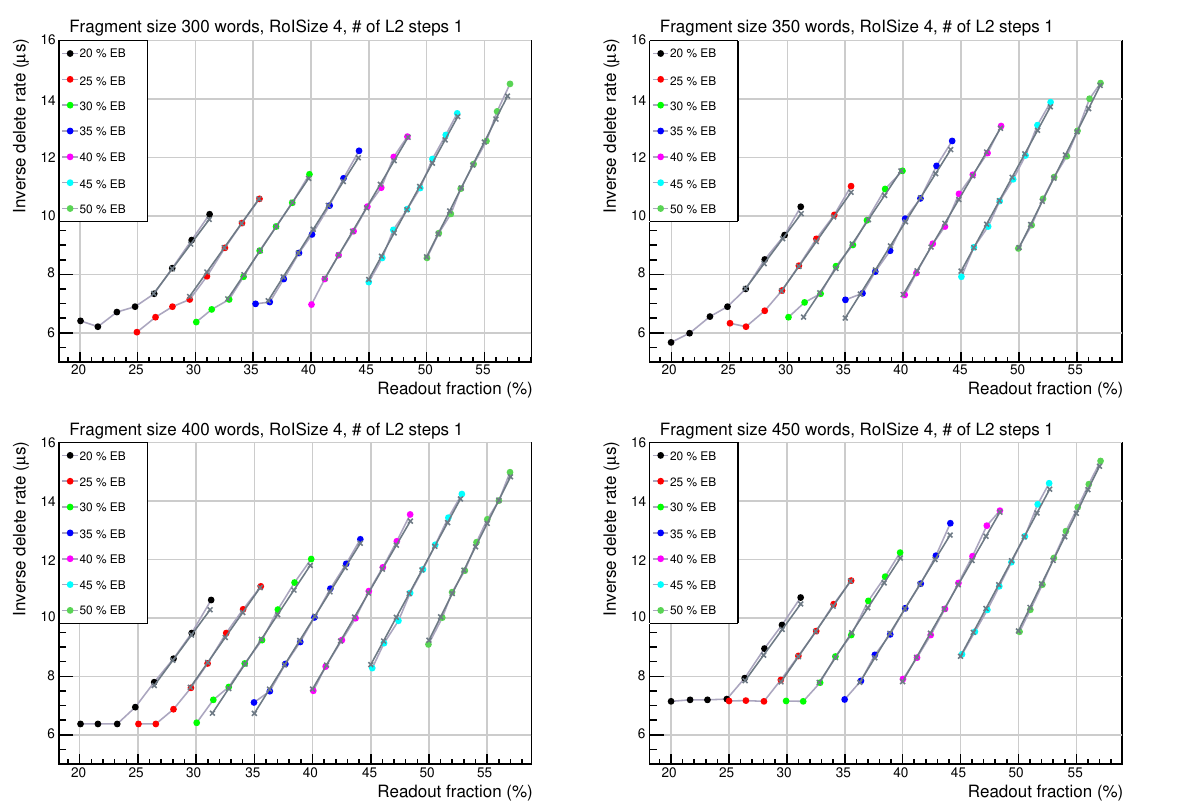}
	\caption{The inverse delete rate (effectively the average time between deletes) observed as a function of the readout fraction for the EB request fractions and fragment sizes indicated, for an RoI size (R) of 4. The coloured dots represent the measurement results and the grey crosses results for the same readout fractions, calculated from a fit to these and other measurement results with the measured inverse delete rate required to be larger than 8~$\mu$s for a fragment size of 450 words and larger than 7~$\mu$s for the other fragment sizes. The lines connecting the dots and crosses have been drawn to guide the eye.}
		\label{fig:invDelRate}
\end{figure}

Figure~\ref{fig:invDelRate} shows the inverse delete rate (i.e. effectively the average time between deletes) as a function of the overall readout fraction for 300, 350, 400 and 450 word fragments, an RoI size of 4 and for EB request fractions in the range from 20 to 50\%. This quantity will be used throughout this section as the linear dependence between it and the overall readout fraction are useful for modelling. The overall readout fraction is itself defined as the fraction of the input data that is requested, calculated from the EB and L2 request fractions, $f_{EB}$ and $f_{L2}$ as follows:

\begin{equation}
f_{readout} = \frac{f_{EB} * N - f_{EB} * f_{L2} * R  + f_{L2} * R}{N}
\label{eqn:readoutfraction}
\end{equation}

where $N$ is equal to the number of input links (24 for a ROS server with two RobinNP cards) and $R$ is the RoI size. Data requested with an L2 request followed by an EB request associated with the same L1ID, not requested again by the EB request, gives rise to the term $ f_{EB} * f_{L2} * R$. 

The results in Figure~\ref{fig:invDelRate} show the effect of the 250 MB/s bandwidth limitation of the internal firmware data generator. As the fragments include a 13 word header added during processing the bandwidth limitation implies a maximum delete rate of 143 kHz for 450 word fragments and of 161 kHz for 400 word fragments, i.e. a minimum inverse delete rate of 7.0 $\mu$ and of 6.2 $\mu$ respectively. For 300 and 350 word fragments and inverse delete rates below 7 $\mu$s the rate appears to be limited by factors other than the 250 MB/s bandwidth. For higher request fractions an approximately linear dependence between delete rate and readout fraction is observed. This dependence can be attributed to data processing taking place in the ROS server, with the linearity of the relationship suggesting that the effective processing time, in which the effect of parallel processing is absorbed,  can be assumed to be a linear function of the L2 and EB request fractions, the delete rate and of the fragment and RoI sizes. The total effective processing time will depend on the rate with which data requests and delete requests arrive, how often they result in internal data transfers (i.e. DMA transfers from RobinNP memory to server memory) and the amount of data to be handled in each case. This dependence is modelled by associating with each type of request (EB, L2  or delete) and with the transfer of event data an effective processing time. The sum of the products of the request frequencies and effective processing times for each request type and of the product of the total amount of data transferred in kB and the effective processing time associated with this transfer should then be 1 in the case of 100\% utilisation of the ROS server's processing resources. This therefore gives a useful indication of the overall load on a system, which can be expressed as:

\begin{equation}
U = F_{e} * t_{e} + F_{i} * t_{i} + F_{d} * t_{d} + S *  t_{1k}
\label{eqn:utilisation}
\end{equation}

with $U$ being the overall utilisation of the system (as a fraction, with maximum value 1); $F_{e}$ being the external data request frequency; $F_{i}$ the internal data request frequency; $F_{d}$ the delete frequency; $t_{e}$, $t_{i}$ and $t_{d}$ the effective processing times associated with the type of request; S the average number of kB transferred per second and $t_{1k}$ the effective processing time associated with transferring 1~kB.

For the case that U is 1, this can be transformed into an expression for the inverse delete rate as a function of the request fraction $f_{EB}$, the readout fraction $f_{readout}$, RoI size $R$, fragment size $s$ and number of input links $N$ using:

\begin{equation}
\begin{array}{c@{}c}
F_{e} = F_{d} * (f_{EB} + f_{L2}) \\
F_{i} = F_{d} * (f_{EB} * N - f_{EB} * f_{L2} * R  + f_{L2} * R) = F_{d} * N * f_{readout} \\
S = s * F_{i} = s * F_{d} * N * f_{readout}
\end{array}
\label{eqn:fractions}
\end{equation}

Combining \ref{eqn:fractions} and \ref{eqn:utilisation} with U set to 1 gives:
\begin{equation}
\frac{1}{F_{d}}= (f_{EB} + N * (f_{readout} - f_{EB}) / (R * (1 - f_{EB})) * t_{e} + N * f_{readout} * t_{i} + t_{d} +  s * N * f_{readout} * t_{1k}
\label{eqn:inverserate}
\end{equation}

This shows that, for this model with fixed $f_{EB}$ and for 100\% utilisation, the inverse delete rate depends linearly on $f_{readout}$.\footnote{$f_{readout}$ was introduced to facilitate the discussion of the measurement results. However, this is not strictly necessary, by combining equations~\ref{eqn:fractions} and~\ref{eqn:inverserate} with U set to 1 the minimum inverse delete rate can also be expressed in terms of the L2 and EB request fractions only:

\begin{equation}
\frac{1}{F_{d}} = (f_{EB} + f_{L2}) * t_{e}  + ( t_{i} +  s* t_{1k}) * (f_{EB} * N - f_{EB} * f_{L2} * R  + f_{L2} * R) + t_{d}
\end{equation}
}

The model has 4 parameters: $t_{e}$, $t_{i}$, $t_{d}$ and $t_{1k}$. By fitting to the measurement results above, the following values were obtained:

\begin{equation}
\begin{array}{c@{}c}
t_{e} = 5.7~\mu s \\
t_{i} = 0.24~\mu s \\
t_{d}  = 0.89~\mu s \\
t_{1k} = 0.13~\mu s / kB
\end{array}
\label{eqn:utilisation_params}
\end{equation}

These can be considered to be the key quantities necessary to model the performance of a given hardware configuration. With this performance signature the overall utilisation of a given ROS server can be computed if one measures the request rates and bandwidths required to complete equation~\ref{eqn:utilisation} for the operational scenario being studied.


Looking in more detail at the results, the model shows that requesting L2 data in more than one step can have a significant impact on the maximum achievable delete rate. This is due to the increase of the external request rate and the relatively large value of the effective time associated with handling an external request. For example for a fragment size of 400 words, EB fraction of 20\%, L2 fraction of 70\% and an RoI size of 4 the model predicts a maximum delete rate of 108~kHz, while when requesting the same RoI data with L2 requests for an RoI size of 2, i.e. with twice the L2 request rate and therefore an L2 fraction of 140\%, this rate drops to 75~kHz. For this case replacing all L2 requests with EB requests, given that L2 requests may cause subsequent requests for the rest of the data for the same L1ID, would result in $f_{EB} = 76\%$ (i.e. 70\% plus 20\% of the 30\% of events for which no L2 request but are instead subject to an EB request) and a comparable maximum delete rate of 74 kHz. However, replacing L2 requests with EB requests results in higher throughput across the PCIe interfaces and network. The total PCIe bandwidth available is about 3.2 GB/s for two RobinNPs. Along with the fragment size, this determines the maximum readout fraction for a given delete rate, assuming not all processing capacity is used. With four 10 GbE links the total available network bandwidth is larger than the PCIe bandwidth, assuming network redundancy constraints don't provide a smaller limit.

\subsection{Performance During ATLAS Data Taking}
\label{subsec:run2_perf}

Across Run~2, operational data for the readout system were routinely recorded in all ATLAS data taking sessions. This has provided a wealth of information with which to assess the performance of the updated ROS as well as to identify areas where the system could be improved for Run~3.
Overall the performance of the upgraded ROS comfortably satisfied the ATLAS requirements for Run~2, with significant headroom within which to accommodate evolving conditions. As will be discussed later in this section, performance limitations only began to emerge at the end of Run~2 when the system was used beyond design specifications in high pileup runs with an HLT configuration generating a large number of L2 requests per event. 

\begin{figure}[h]
	\centering
	\begin{minipage}{0.32\textwidth}
		\includegraphics[width=\textwidth]{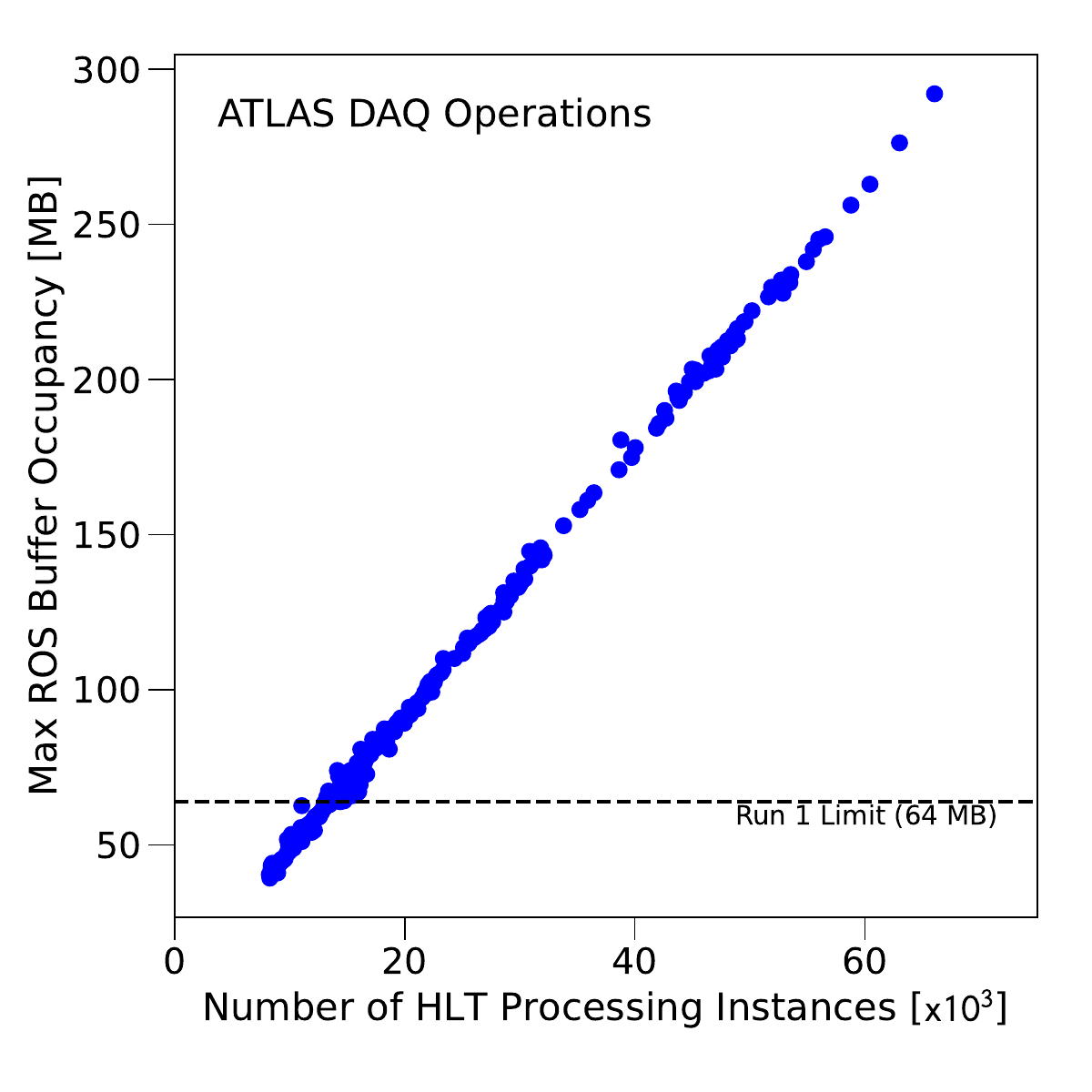}
	\end{minipage}
	\begin{minipage}{0.32\textwidth}
		\includegraphics[width=\textwidth]{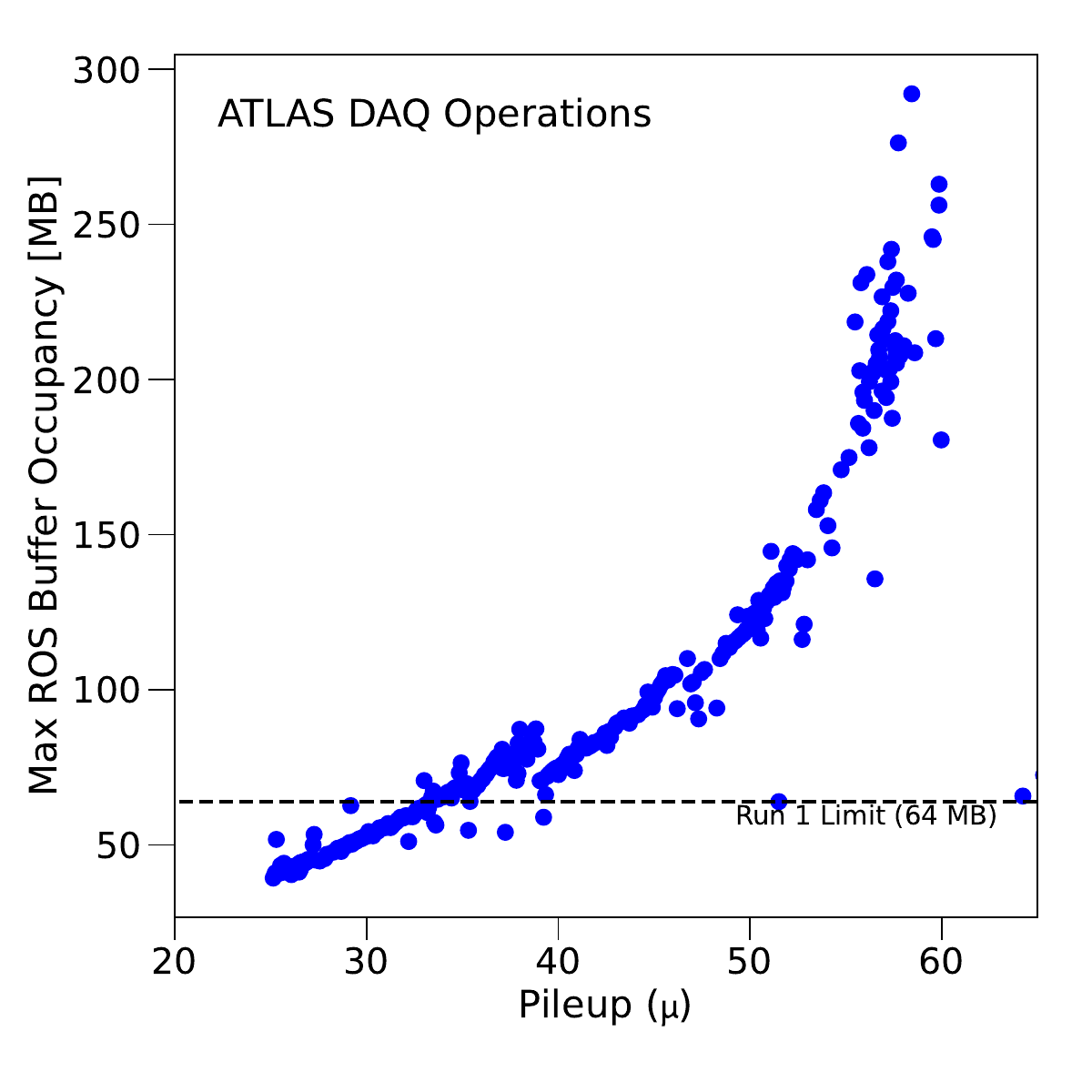}
	\end{minipage}
	\begin{minipage}{0.32\textwidth}
		\includegraphics[width=\textwidth]{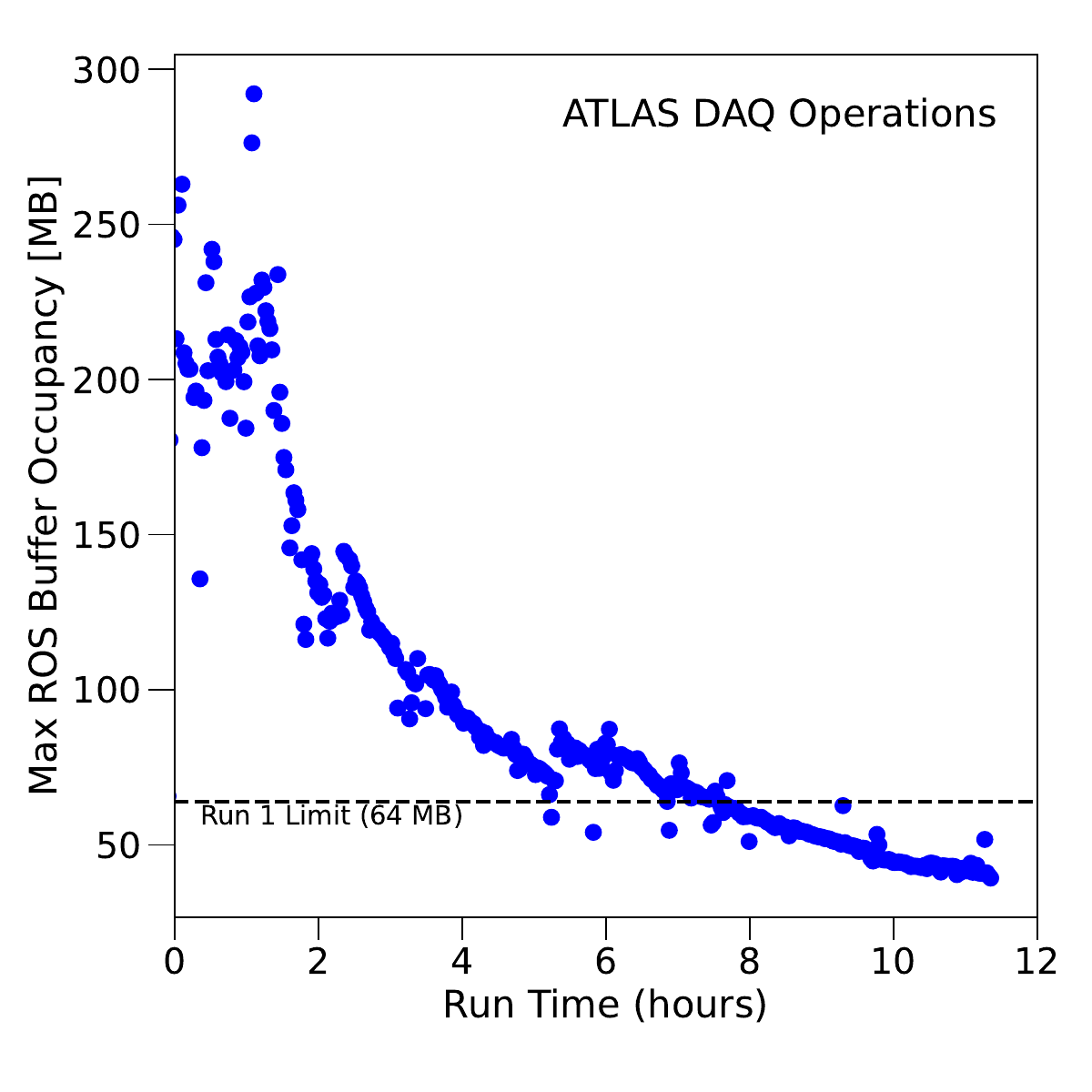}
	\end{minipage}
	\caption{\label{fig:rosperf}
		Performance plots based on operational data taken from an ATLAS run from October 2017, with a peak luminosity at start of run of 1.7$\times10^{34}\textrm{cm}^{-2}\textrm{s}^{-1}$ and peak average pileup of 66.5. Such luminosity and pileup conditions are expected to be replicated in Run~3. (left) Max ROS buffer occupancy against the number of HLT application instances in use in the HLT farm.  (middle) Buffer occupancy versus pileup. (right) The evolution of max buffer occupancy over time across the run. The dashed line indicates the 64 MB of buffer space available per ROL for Run 1.}
\end{figure}

To demonstrate how the updated ROS allowed ATLAS to exceed the limitations that would have existed had the Run~1 ROS continued to be used, Figure~\ref{fig:rosperf} shows the impact of the increased per-channel memory capacity of the RobinNP by comparing the limits imposed by the Run~1 ROS on HLT farm size (left), as well as the evolution of buffer occupancy throughout a run (middle) and the dependence on pileup (right). In all three cases the single readout channel with the highest buffer occupancy across all systems is shown for the corresponding number of HLT processing instances, run time and pileup respectively. The Run~1 ROS had a buffer capacity of 64~MB per channel, whereas for the Run~2 ROS this increased to 670~MB.

The size of the ROS buffer limits the HLT farm size because it places a constraint on the average time an HLT node can take to process an event (i.e. the latency). Given a fixed input rate to the farm of 100~kHz, it is desirable to increase the number of available nodes such that this latency value can grow, meaning more sophisticated trigger algorithms can be executed. A small ROS buffer would cap this growth, thus limiting the scope for algorithmic complexity. This is shown in the left-hand figure, where it can be seen that the Run~1 limitation of 64 MB would have limited the farm size to approximately 15,000 HLT application instances, compared to the approximately 70,000 active in Run~2 (with the capacity for this to increase further). 

Looked at from another perspective: the high pileup conditions at the start of an LHC run mean events take the longest to process in the HLT. Were it not possible to allow sufficient latency budget for the trigger algorithms to operate then the configuration of the trigger would have to be adapted, either to produce a lower input rate at L1 or with a less complex and physics rich set of algorithms in the HLT. The middle figure shows this would be the case at the start of many ATLAS runs. Finally, looking specifically at time, the right hand figure shows that, were it not for the updated ROS, the overall system occupancy would have rendered the trigger configuration unviable for a significant fraction of typical ATLAS runs in Run~2.

\begin{figure}[h]
	\centering
		\begin{minipage}{0.32\textwidth}
			\includegraphics[width=\textwidth]{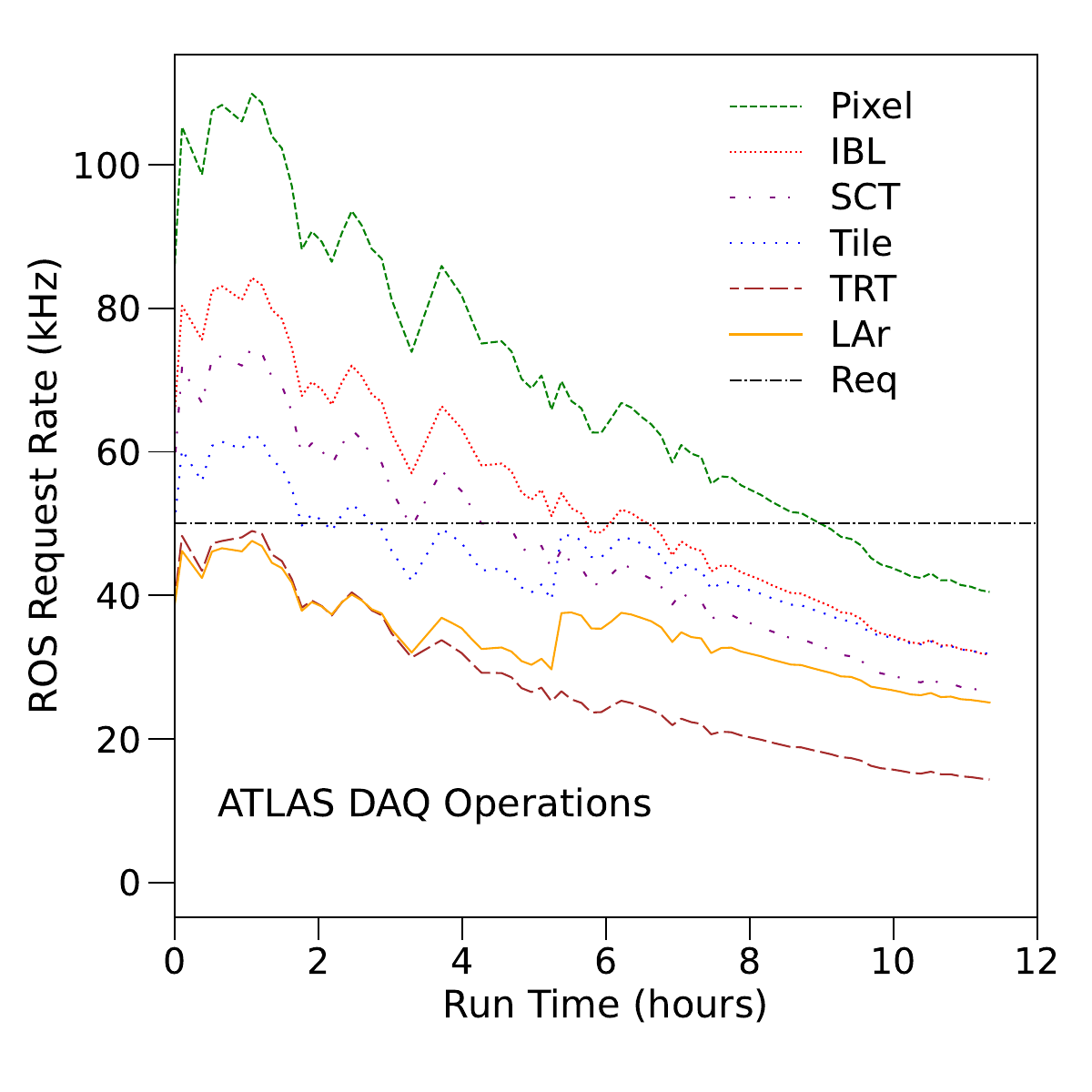}
		\end{minipage}
		\begin{minipage}{0.32\textwidth}
			\includegraphics[width=\textwidth]{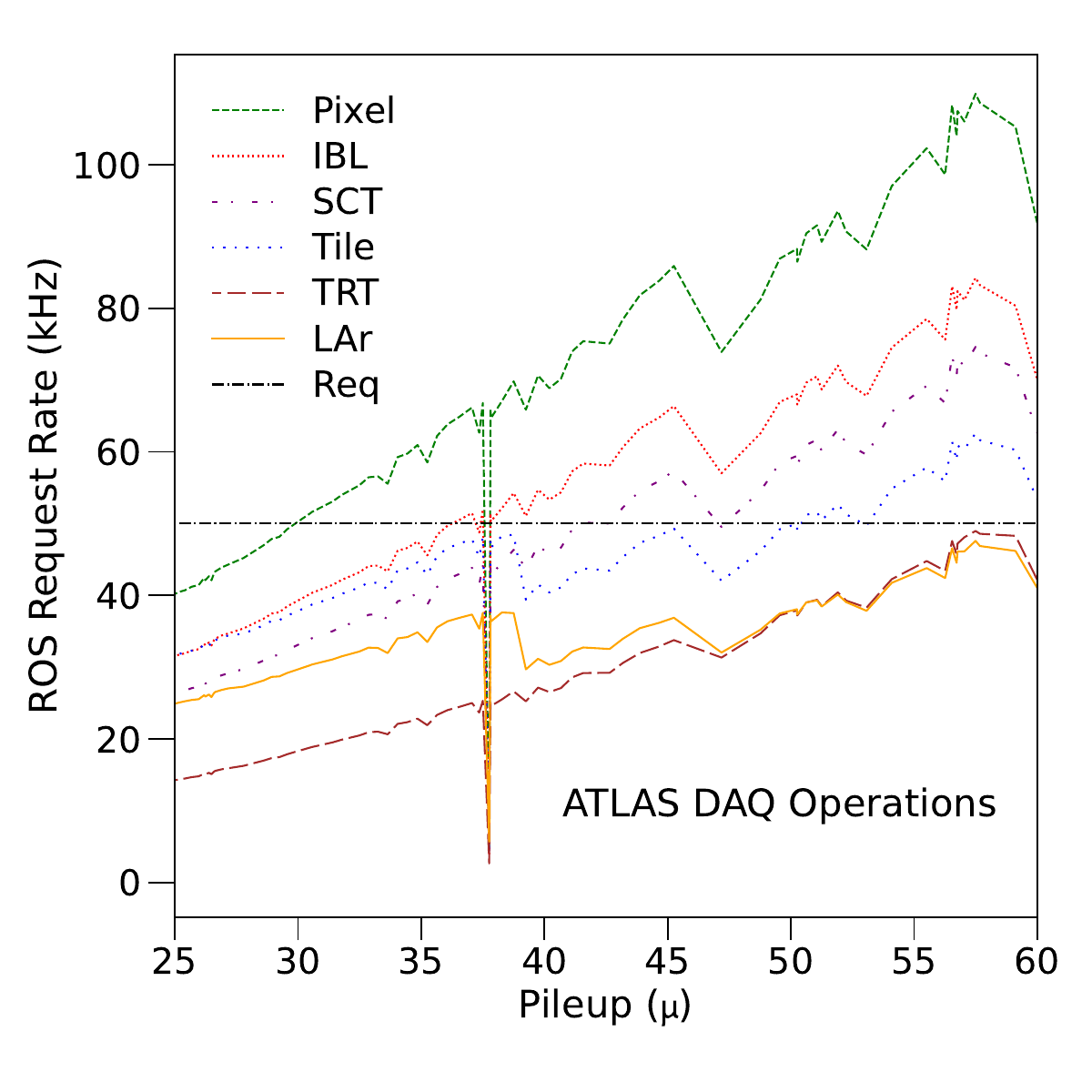}
		\end{minipage}
		\begin{minipage}{0.32\textwidth}
			\includegraphics[width=\textwidth]{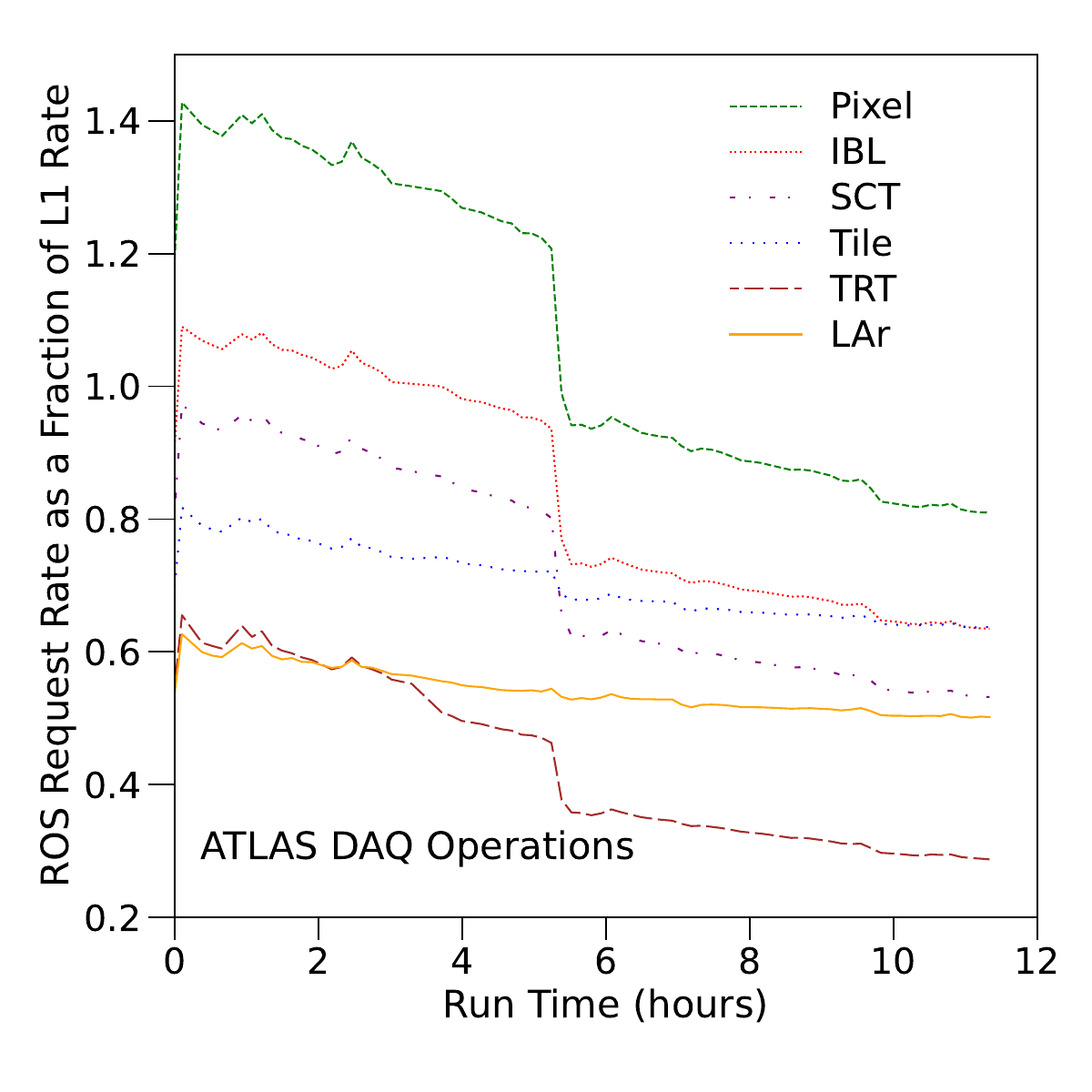}
		\end{minipage}
\caption{\label{fig:p1reqrates} (left) Maximum request rates for the most heavily loaded ROS systems for particular detectors during an ATLAS run from October 2017, with a peak luminosity at start of run of 1.7$\times10^{34}\textrm{cm}^{-2}\textrm{s}^{-1}$ and peak average pileup of 66.5. (centre) Maximum request rates for the same ROS servers in the same run, this time as a function of pileup. The rates linearly depend on the pileup, with the Pixel detector most frequently requested across the range. In both left and centre plots the maximum expected request rate (Section~\ref{subsec:requirements}) is shown as a horizontal dot-dashed line labelled 'Req'. The discontinuity near pileup of 38 is due to a temporary pause in data taking during the run and can be ignored. (right) Maximum request rates as a fraction of L1 rate as for the same ROS servers in the same run, once again as a function of time. As can be seen, sections of the Pixel detector readout receive requests at a rate significantly above the L1 rate for the first half of the run. The sharp changes in rate at 5.5 hours (left) and at a pileup of 38 and 47 (right) were due to prescale changes made during the run to optimise HLT farm occupancy, accounting for the fact that the L1 rate will otherwise decay over time. In all plots the order of entries in the legend match the order of appearance the corresponding lines. }
\end{figure}

As a more direct measurement of the performance of the ROS, Figure~\ref{fig:p1reqrates} shows the observed readout request rates for selected heavily loaded ROS servers for the same run as Figure~\ref{fig:rosperf}. As can be seen, the rates significantly exceed the original specification of the system in Section~\ref{subsec:requirements} for some of the selected systems (most prominently the Pixel detector) in the early parts of the run (when pileup can be expected to be highest). In the early portion of the run, the trigger configuration is also seen to be generating rates of requests to the ROS larger than the L1 accept rate, meaning that the system in question is responding to more than one request per L1 Accept.

\subsubsection{Observed Performance Limitations During Run~2 and Run~3 Projections}
\label{subsubsec:limitations}

While not directly related to the design of the ROS, an important performance limitation in terms of achievable L1 trigger rate was the occupancy of the S-LINK channels bringing data into the system from the subdetector RODs. For some detectors, such as the tracking detectors, the fragment size produced for each event depends strongly on the collision environment. As the luminosity and pileup increased throughout Run~2, some reached a situation where the standard S-LINK bandwidth (160 MB/s) was insufficient for them to transfer their fragments to the ROS at the required 100~kHz rate.

\begin{figure}[h]
	\centering
\includegraphics[width=\textwidth]{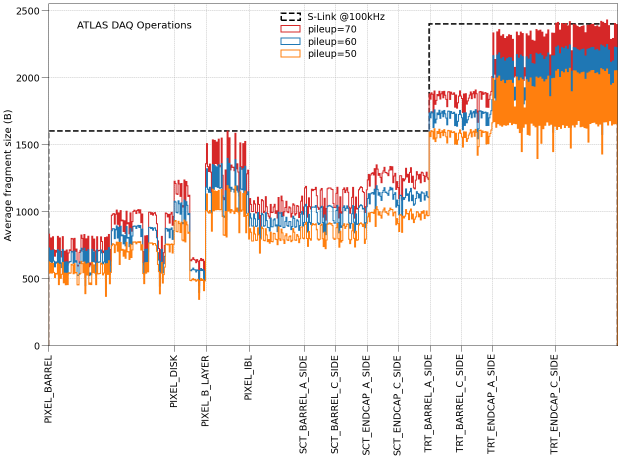}
\caption{\label{fig:link_occupancy1}  {Distributions of fragment sizes for different pileup regimes for readout slices from the inner tracker (pixels, SCT andt TRT) nearing the limit of S-LINK occupancy. Data are obtained by scaling the distribution from an ATLAS run in mid 2018 with peak average pileup of 54.9 and peak instantaneous luminosity of 2$\times10^{34}\textrm{cm}^{-2}\textrm{s}^{-1}$}. The maximum occupancy of the S-LINK implementation for the given systems at 100 kHz L1 rate is shown by the horizontal dashed line. The data series for a pileup value of 70 corresponds to the highest average fragment size for each readout slice, with the series for a pileup value of 50 corresponding to the lowest fragment size and the series for a pileup value of 60 in-between. As can be seen, for parts of the Pixel and TRT detectors the average fragment size is such that it will not be possible to operate at a 100 kHz in very high pileup conditions without further adaptations. For all systems not included in this plot the distributions show that significant margin remains before the S-LINK occupancy limit and thus no further action is required. Variations in fragment size within a given detector type are typically explained by differences in geometrical acceptance.}
\end{figure}

To mitigate this limitation, numerous modifications were made. In some cases, it was possible to modify the configuration of the front-end hardware (ROD) to operate the S-LINK at higher bandwidths (up to 250 MB/s) whereas for others more use was made of data compression or the splitting of a given link such that the same payload is instead routed to two or more links to the ROS. Such updates were not possible for all systems due to front-end hardware constraints. The fragment size for subdetectors which remain close to the operation limit is shown in Figure~\ref{fig:link_occupancy1}. As can be seen, for some detectors, the highest pileup conditions result in sizes which surpass the ability of their S-LINK bandwidth to transfer the data at 100~kHz. Such extreme pileup values are not expected to be common in Run~3, but the situation will nonetheless be monitored in case further mitigation is needed.

Towards the end of Run~2, a new performance limitation began to emerge as the request rates from the HLT to specific ROS servers increased. The increase was due to an interplay between the collision environment and the configuration of the trigger itself, with newer algorithms making an increasing number of requests from the system. As a result, in some runs ROS request rate saturation was observed, with the system unable to service requests at a sufficiently high rate. This was mitigated in the first instance by re-optimising the trigger algorithms and the request patterns sent to the ROS. For example, in some scenarios it was decided it would be better to 'pre-fetch' all data for part of an event (driven by information from the RoI) rather than send multiple requests (for background on why this is useful see the discussion in Section~\ref{subsubsec:utilisation}). Though not always the best approach, strategic deployment of this mode was able to mitigate some of the bottlenecks. 

\begin{figure}
	\centering
	\includegraphics[width=1.0\linewidth, trim={1cm 1cm 0 0},clip]{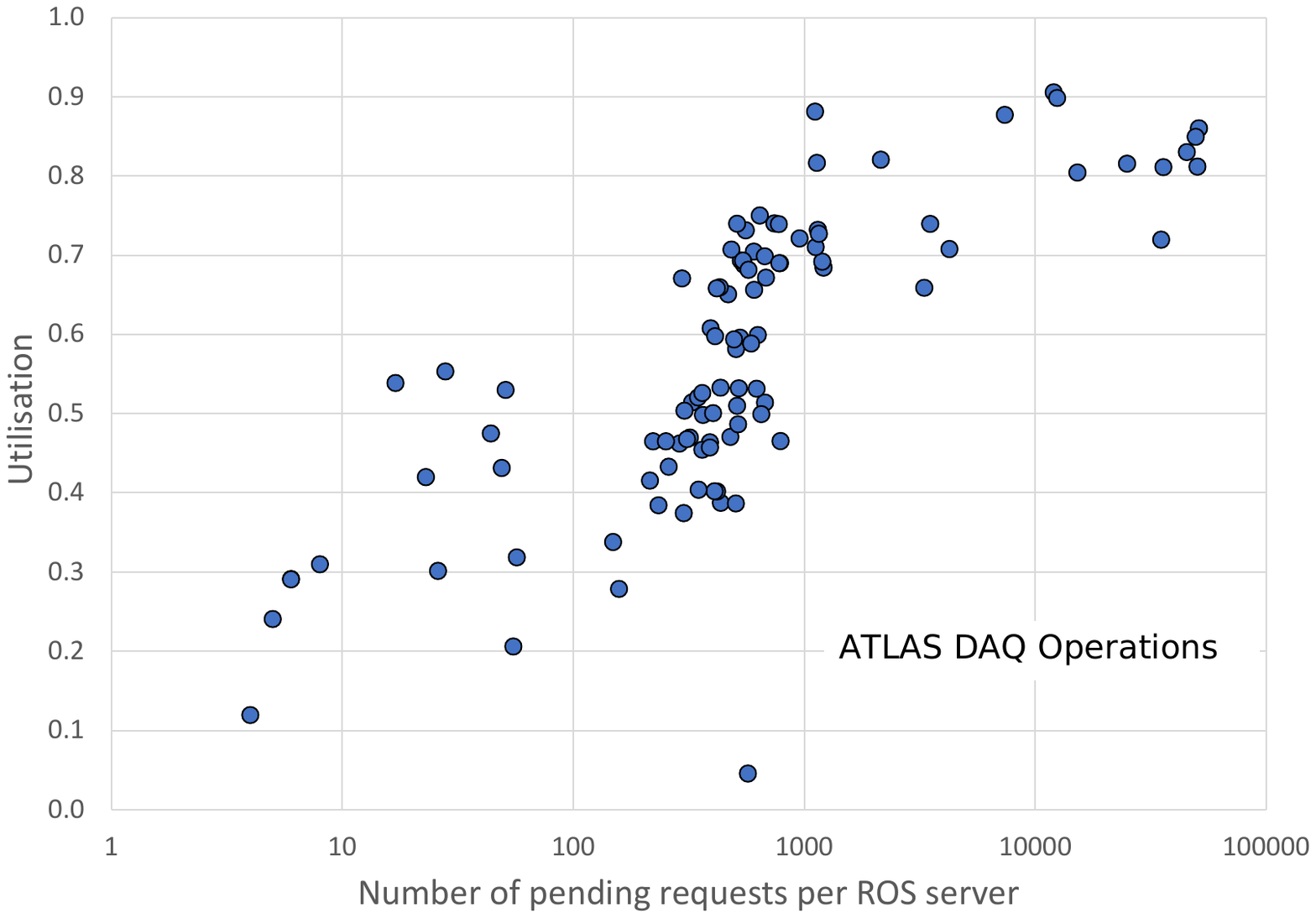}
	\caption{Estimated ROS utilisation (see equation~\ref{eqn:utilisation}) for individual servers plotted against maximum number of pending requests per ROS server for a 1 hour window during an ATLAS run from August 2018, during which a performance limitation was experienced. A number of pending requests greater than 1000 was associated operationally with the system reaching saturation.}
		\label{fig:rosutilisation}
\end{figure}

To better compare system performance during Run~2 with the expectation from laboratory testing, equation~\ref{eqn:utilisation} can be used to quantify whether or not a given ROS server is operating at near its full capacity. By using the reference period values from equation~\ref{eqn:utilisation_params}, along with request frequencies and bandwidths measured from real operational monitoring data, it is possible to compute an estimate of the utilisation value for a ROS server. Figure~\ref{fig:rosutilisation} shows the result of performing this calculation for all ROS servers at a time during an ATLAS data taking session where the server performance limitations were observed. An important indicator that a ROS server is close to being unable to keep up with the request rate is the number of pending requests counted by the monitoring system. By plotting the computed utilisation against the maximum number of pending requests it is possible to show that several servers with a large number of pending requests were indeed operating at between 80-90\% of their capacity, which can be expected to give rise to the observed queueing. This observation confirms both that the laboratory performance benchmarks and formulae can be used to model and predict the performance of the system in ATLAS data taking, but also that the limitations observed did constitute the Run~2 hardware genuinely reaching its operational ceiling.

\begin{figure}[h]
	\centering
\includegraphics[width=1.0\linewidth]{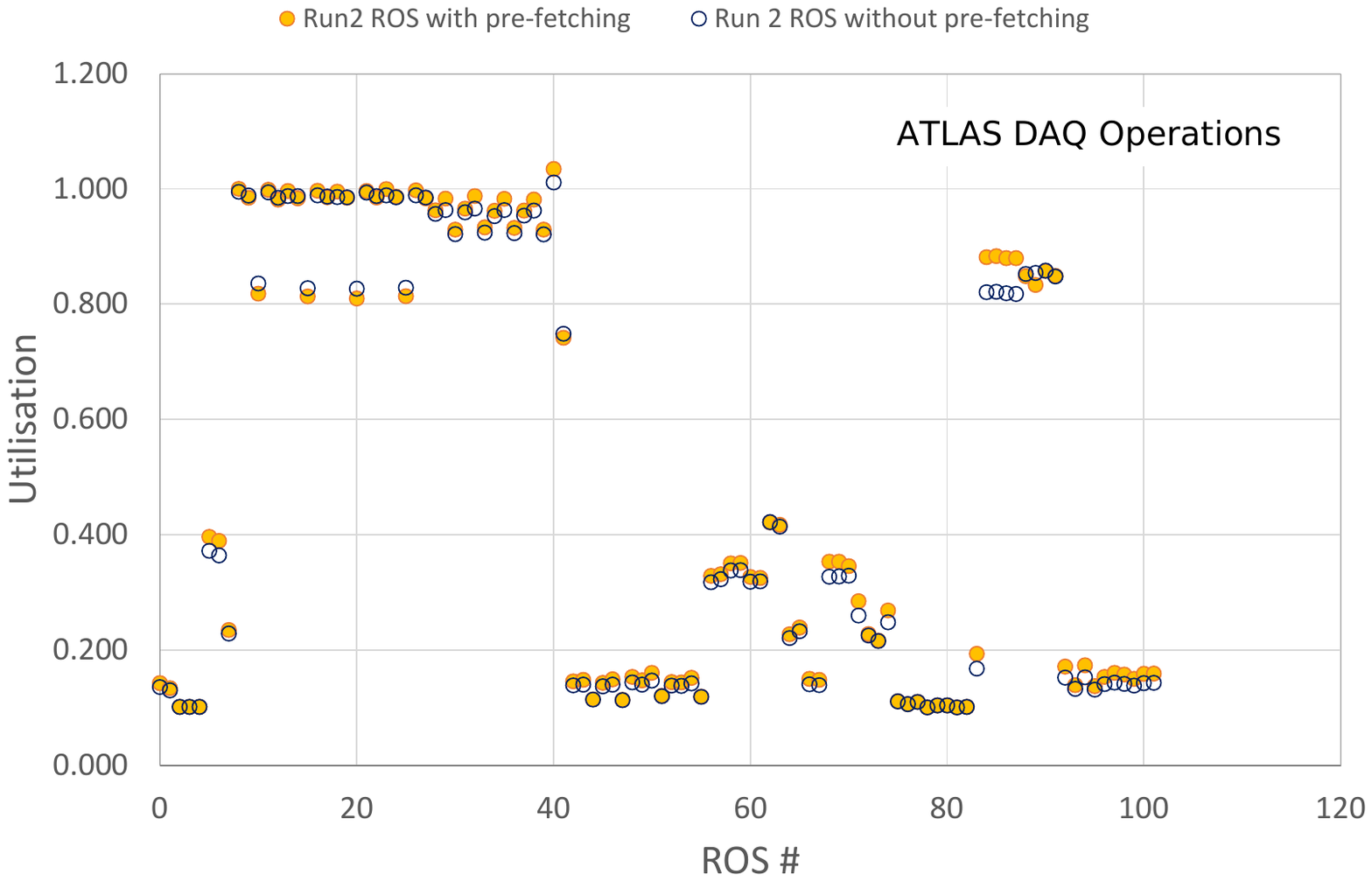}
\caption{\label{fig:l2_saturation} {Predicted utilisation (see equation~\ref{eqn:utilisation}) for Run~2 ROS server hardware based on estimated request rates for Run~3 Trigger configuration, assuming similar luminosity and pileup conditions as Run~2, but for updated versions of server operating system and TDAQ software that were deployed during Run~2. The X-axis shows all servers in the system enumerated from 0 to 101 and grouped by detector system. The maximum utilisation above which saturation is typically experienced is estimated to be in the region of 0.8. For each ROS two points are shown. One for the nominal configuration (i.e. without pre-fetching) and the other with pre-fetching enabled.  As can be seen, for two groups of servers the predicted request rates significantly exceed the 0.8 limit in both pre-fetch and non-pre-fetch regimes. The left hand group with the largest utilisation values corresponds to the Liquid Argon (LAr) Calorimeter and the right hand group corresponds to the Tile Calorimeter. In many cases the pre-fetch case has a higher utilisation value than the non-pre-fetch case. This is expected to be due to the pre-fetch implementation proposed for the start of Run~3 requesting more of the data at once for each event than in Run~2 (see main text for a more detailed description).}}
\end{figure}

Given the above, it was clear that these limitations would likely be experienced once again in Run~3, as there was no longer significant margin left to allow for further trigger developments and to match the evolution of the collision environment. To better model the likely impact of the planned trigger configuration in Run 3, information from the Trigger group on expected request rates per ROS server was used to extrapolate utilisation values based on the Run 2 hardware parameters calculated above. Trigger rates were estimated assuming the same luminosity and pileup conditions as Run~2, which are expected to remain broadly constant in Run~3. The results are shown in Figure~\ref{fig:l2_saturation}, where it is clear that while for many systems the Run~2 hardware would provide sufficient performance, for the calorimeters there is a significant risk of ROS saturation causing performance bottlenecks. It should be noted that the pre-fetching in use in these predictions is different to that from Run~2. Previously, pre-fetching was RoI-driven, meaning a subset of the data for a given event are requested, but in a way that still provides some aggregation of requests compared to the non-pre-fetched case. In Run~3, the initial baseline pre-fetching mechanism was to use 'full' pre-fetching, where all data for a given event are requested. This was replaced soon after the start of Run~3 by something closer to the Run~2 equivalent, but the full pre-fetch mode was used to generate the predictions shown in the figure. This likely explains why the pre-fetch performance is in some cases worse than without it.

Based on these results, it has been decided to adopt a two-fold strategy to mitigate the problem. First, a small number of additional ROS servers have been added to share the load of the most heavily used systems (primarily in the Liquid Argon Calorimeter Hadronic Endcap). In this case, the links serviced by one heavily loaded server have been split between two. Second, given they are ageing and no longer protected by warranty, the servers and network interface cards themselves have been replaced with newer models (with the RobinNPs unchanged) at the time this paper was finalised, giving the opportunity to benefit from performance advances in the six years since the system was first commissioned. This evolution will be described in more detail in Section~\ref{sec:run3}.

\subsection{System Reliability}
\label{subsec:reliability}

The updated ROS has proven extremely reliable during Run 2. From a hardware perspective, less than 1\% of RobinNP cards and ROS servers experienced any kind of malfunction impacting on data taking during the entire run period. Where problems did occur they were typically recoverable (e.g. fan or DIMM replacement). From a firmware and software perspective, the RobinNP and ROS contributed a negligible amount to overall ATLAS DAQ downtime during the same period.

\clearpage
\section{ROS Technology re-use in ATLAS}
\label{sec:techreuse}

The RobinNP and upgraded ROS development have provided some good examples of technology re-use. First and foremost the RobinNP itself is implemented on hardware designed for a similar role by the ALICE Collaboration~\cite{Engel:2683612}. While the firmware is ATLAS-specific, the end stages of hardware design, as well as the bulk fabrication process, were a joint enterprise~\cite{borga}. Such a collaboration made it possible to reduce costs while also exposing the designs on both sides to a wider pool of expertise with mutual benefits. Within ATLAS the ROS has also proved to be a flexible and re-usable technology. In this section a description will be given of three examples of this effect, the first of which has provided significant operational benefit to ATLAS.

\subsection{Region-of-Interest Builder (RoIB)}
\label{sec:pc_roib}
The PC-RoIB uses the new ROS technology for the replacement of the ATLAS Region-of-Interest Builder (RoIB). Previously this unit was implemented using custom VMEbus modules~\cite{Blair:2008}. The RoIB receives fragments over S-LINK from various parts of the L1 trigger, assembles the fragments from a single collision into a record which is then forwarded to the HLT farm supervisor that assigns each event to a free HLT node for processing. Whilst the performance required was beyond that of the original ROS, the performance of the new ROS technology is sufficient to handle the task. During the 2015-2016 winter shutdown the RoIB was upgraded to the new ROS PC based design~\cite{Abbott:2016}, with a dedicated software API for Region-of-Interest building, allowing it to be integrated with the existing farm supervisor process (HLTSV), running on the same ROS server as the RoIB software. The upgrade to ROS technology has eliminated the need for some ageing custom hardware for which maintenance was becoming an increasing concern. The resulting performance and monitoring improvements allow for further scaling to cover future changes to requirements.

\begin{figure}[hbtp!]
    \begin{centering}
    \includegraphics[width=0.8\linewidth]{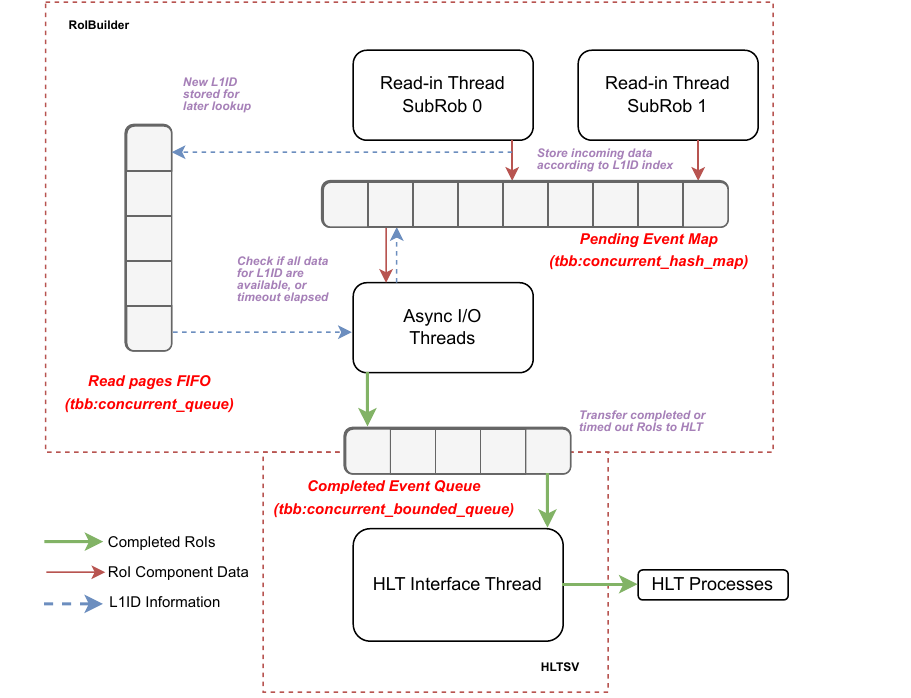}
    \caption{Schematic of the multi-threaded processes of the PC-RoIB and the memory structures that are accessed.}
    \label{fig:PCRoIBArch}
    \end{centering}
\end{figure}

The main functionality of the PC-RoIB is implemented in four threads using TBB~\cite{intel:tbb} data structures to cross between thread domains.  Figure~\ref{fig:PCRoIBArch} shows the schematic of the multi-threaded application and the memory structures. Two dedicated SubRob-specific threads read the data triggered DMA queues and put the data into a map of pending events indexed by the L1ID.  For the first fragment the L1ID is also sent to the queue of the read pages FIFO.  The Result Handling thread monitors the read pages FIFO using a polling loop to check if for any L1ID in the system all fragments from the active channels have been received or the event time out threshold have been reached. If this is the case the fragments are transferred from the input buffers to the completed event queue. A dedicated thread interfaces to the existing HLTSV software and returns the most recent L1ID and frees the DMA buffers into which the data were originally transferred by the RobinNP.  The DMA pages are kept in the system to allow the RoIB to assert flow control to the upstream system in the event that there are no slots in the CPU trigger farm for the HLTSV to send a L1ID. During normal operation when no flow control is necessary the latency of the PC-RoIB software is 2.8~$\mu$s on average.

\subsection{QuestNP}

The QuestNP is an S-LINK data source card based on C-RORC hardware, implemented as a new firmware design re-using elements from the firmware of the RobinNP and the conceptual design of the original QUEST~\cite{Haas:796120}. The QuestNP makes it possible for user defined data to be transferred to a remote S-LINK data receiver in parallel on each of the 12 S-LINKs supported by the board at a fragment rate of 100 kHz (assuming the fragment size is not so large as to cause link saturation). The primary client driving QuestNP was the ATLAS Fast TracKer (FTK)~\cite{ftk}, for whom the board was used to facilitate system-level commissioning with realistic data. Given that the data payload to transfer is fully user-controlled, it is also possible to send fragments with known problems in order to assess their effect on the receiving hardware. This ability to measure the impact of known problematic data was considered vital to improve the robustness of FTK.

\subsubsection{Firmware Design}

\begin{figure}[h]
	\begin{centering}
\includegraphics[width=\textwidth]{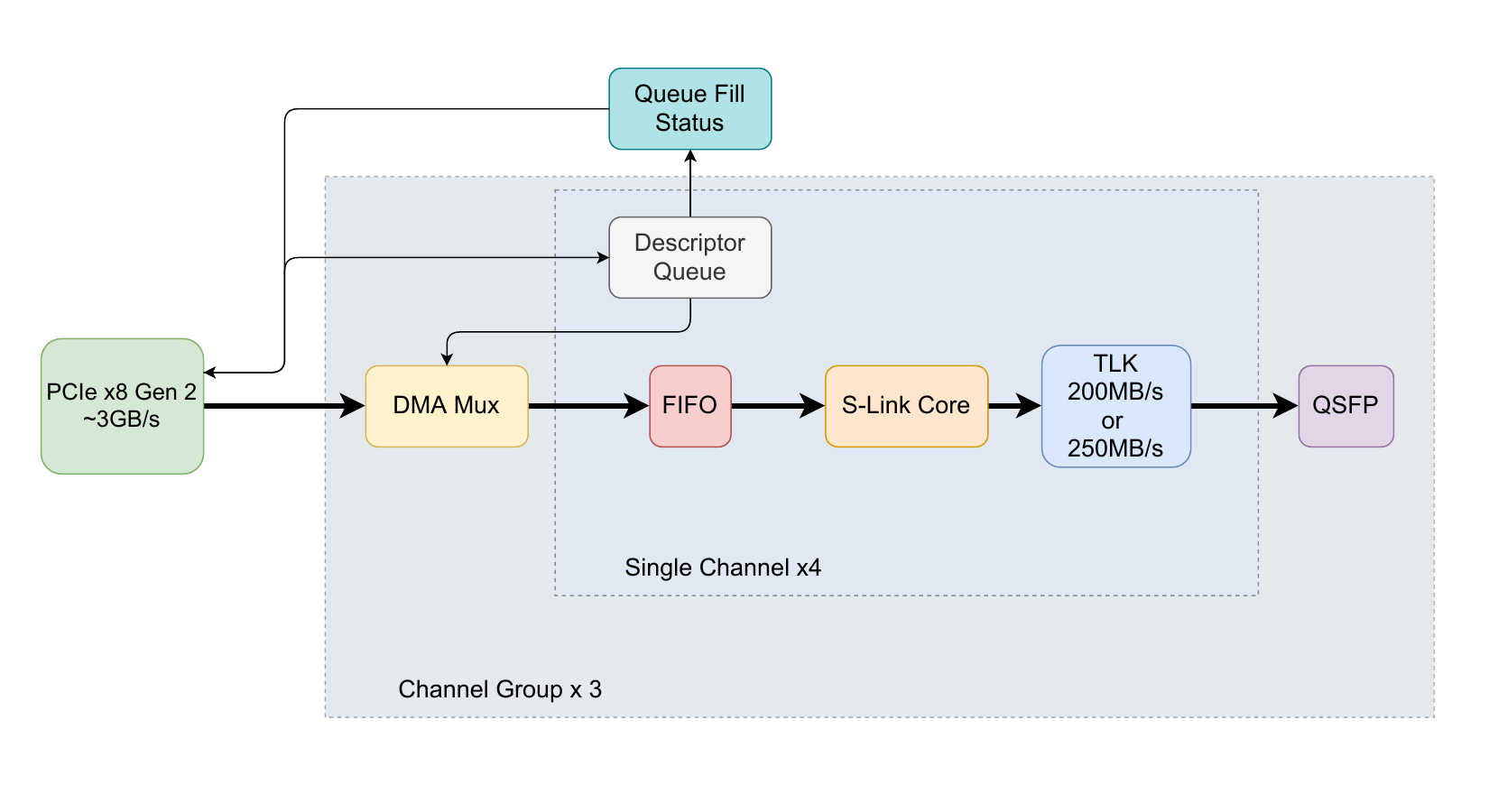}
\caption{\label{fig:questnpdma} Diagram of the logical firmware blocks making up the QuestNP. The PCIe interface and DMA engine, as well as the S-LINK and TLK blocks were re-used from the RobinNP codebase.}
    \end{centering}
\end{figure}

The QuestNP firmware design is shown in Figure~\ref{fig:questnpdma}. The 12 output channels of the board are split between three channel groups, each servicing four channels. Each channel group has a DMA demultiplexer, through which transfers from host memory are routed to the appropriate channel. Each channel then has its own S-LINK wrapper and internal buffering. The PCIe interface and DMA engine, as well as the S-LINK and TLK blocks were re-used from the RobinNP. Given that the QuestNP does not make use of the on-board memory buffers provided by the C-RORC, the overall design is significantly simpler than that for the RobinNP. As such, it was possible to implement an 8 lane PCIe Gen 2 interface, the maximum supported by the C-RORC, giving an expected maximum throughput of 3~GB/s per board.

As well as the required dataflow functionality, the QuestNP firmware also supports the same control and monitoring features over I2C as the RobinNP (see Section~\ref{subsec:mgmt_mon}), meaning the board can be managed using similar (if not identical) tools without additional firmware development. 

\subsubsection{Software Design}

The software package for the QuestNP re-used existing ROS drivers and low level libraries, meaning that additional development was mainly required for higher level tools and user interfaces. Applications were provided to monitor board health and perform low level functionality tests, as well as to service high bandwidth dataflow. For this second scenario, a multi-threaded application integrated with the ATLAS run control infrastructure was provided, which is able to load large user-configurable datasets and stream them via DMA over the PCIe interface for transfer across S-LINK to a remote peer. The thread configuration was then optimised to maximise DMA throughput.

\subsubsection{Potential Future Work}

Should operational need arise, further performance optimisation could be possible by moving to interrupt driven communication when managing DMA requests. It would also be possible to synchronise the L1IDs in the fragments sent across multiple boards by use of a wired connection, operating in a daisy chain, where one channel is nominated as the primary source and all other channels are required to never deviate beyond a given number of events from this ID in flight on this channel.

\subsection{Dozolar}

As an additional testing resource, the C-RORC hardware was also re-used to implement the Dozolar, a 12 channel version of the original DOLAR data source~\cite{Haas:796120}. A simpler design than the QuestNP, the Dozolar does not use DMA transfers to receive data from the host, with a version of the RobinNP firmware data generator (see Section~\ref{subsec:datagen}) having been implemented to instead produce properly formatted fragments at configurable rates to be sent to a peer over S-LINK. The firmware is based on a combination of RobinNP modules and others taken from earlier generation readout boards. The software similarly re-uses drivers and libraries from previous implementations, adding only the features needed to control the rate and service the additional channels. Given all data are generated inside the firmware traffic across the interface is minimal, so the role of the software is purely to control and configure the board. Fragments are generated using the standard ATLAS ROD output format~\cite{Bee:683741} with configurable size (though fixed to that single value) and contain a pre-defined payload pattern. 

The Dozolar made a significant contribution to the testing of the ROS for Run~2. Acting as a remote S-LINK data source it operated stably at the required 100~kHz rate for up to 12 parallel data channels. The board is still regularly in use as part of the validation of firmware and software changes integrated into the ROS.

\clearpage
\section{Run~3 Preparations}
\label{sec:run3}

As discussed in Section~\ref{subsubsec:limitations}, the ROS servers used for ATLAS data taking
have been replaced with new models for Run~3. The RobinNP cards themselves have not been replaced. The primary motivation for the replacement is that the age of the servers (now over seven years of almost continuous operation), is such that their current level of operational reliability cannot be guaranteed through to the end of Run~3, at which point they will be a decade old, with component obsolescence meaning large scale like-for-like replacement isn't feasible. Warranty protection for these servers has also expired, meaning no guaranteed manufacturer support in case of failures.

Given technological evolution since the original installation, it was expected that any newer systems would bring performance improvements which would assist in satisfying the expected Run~3 performance requirements discussed previously. If not, alternatives would have to be considered (such as splitting the link load across multiple servers), resulting in a larger system size overall. As described below, based on the outcome of performance tests a type of server considerably more performant than the Run 2 ROS servers was selected. It has been subject to rigorous stress testing to fully validate it before installation.

\subsection{Selection of Updated Host Server}

In order to select a new server a number of candidates were tested. A key specification is that the new servers should have the same size and network connectivity as those which they will replace, while operating within the same power budget. Specifically, a 2U high box with two redundant power supplies and two dual port 10 GbE NICs. Each server would also have to host up to two RobinNP cards. As such, any suitable motherboard/CPU combination would have to be available in a form factor that fits within these constraints. Based on this, the expected cost and past experience with PCIe lane and memory layout, the decision was made to focus on single CPU servers. From a network perspective, the new server hosts an updated version of the 10 GbE NICs used in Run~2 (Mellanox ConnectX-4~\cite{ROS:MellanoxNICRun3} rather than ConnectX-3). These were selected based on expected operational reliability. Testing in other contexts suggests no significant performance gain is likely to be yielded by an alternative specification. 

After the elimination of other candidates, a small number of systems were chosen for more detailed performance tests. From Intel, a Cascade Lake-era Xeon W based system. From AMD, two Epyc based servers, one based on the older `Rome' generation and another on the more recent `Milan' line. Details of each system tested are presented in Table~\ref{tab:run2_candidates}.

\begin{table}[hbtp!]
\centering
\begin{tabular}{ | l | c | c c c |}
\hline
 Name & Run~2 ROS & Candidate 1 & Candidate 2 & Candidate 3  \\ \hline \hline
 \multirow{3}{*}{CPU} &  Intel E5-1650 V2 & Intel Xeon W-3223  &  AMD Epyc 7302P & AMD Epyc 7313P \\ 
 &  & & (Rome) & (Milan) \\   
 &  6 cores @ 3.5 GHz & 8 cores @ 3.5 GHz & 16 cores @ 3.0 GHz & 16 cores @ 3.0 GHz \\ \hline
 \multirow{2}{*}{RAM} &  16 (4 x 4) GB & 48 (6 x 8) GB  & 64 (8 x 8) GB & 128 (8 x 16) GB\\    
 &  ECC DDR3-1866 &  ECC DDR4-2666 &  ECC DDR4-3200 &  ECC DDR4-3200 \\
 \hline
 NIC &  ConnectX-3 &   ConnectX-3  &   ConnectX-3  &  ConnectX-4 \\
 \hline
\end{tabular}
\caption{\label{tab:run2_candidates} Run~3 ROS server candidate specifications, compared to the Run~2 server on the left. The choice of NIC was shown to not have a significant performance impact, and therefore Candidate 3's results are considered directly comparable to the others.}
\end{table}

By replicating the procedure outlined in Section~\ref{subsubsec:rostester}, it was possible to measure the performance signature of any candidate server for comparison with another for a given request rate and bandwidth scenario. By comparing these results it was concluded that the AMD Epyc `Milan' / Mellanox Connect X-4 system performed best out of all candidates. As a further optimisation an additional study was performed with different RAM DIMM configurations. The baseline (8 x 16 GB) DIMM configuration was compared with one with a single 16 GB DIMM and another with two 16 GB DIMMs (i.e. 32 GB in total). In each case the DIMMs were fitted in slots according to the motherboard manufacturer's recommendations to avoid functional or performance problems. This demonstrated that memory configurations at or above the two-DIMM 32 GB scenario performed very similarly, whereas the the single DIMM case performed significantly worse.

\begin{figure}[hbtp!]
	\centering
	\includegraphics[width=1.0\linewidth,trim={1cm 1cm 0 1cm},clip]{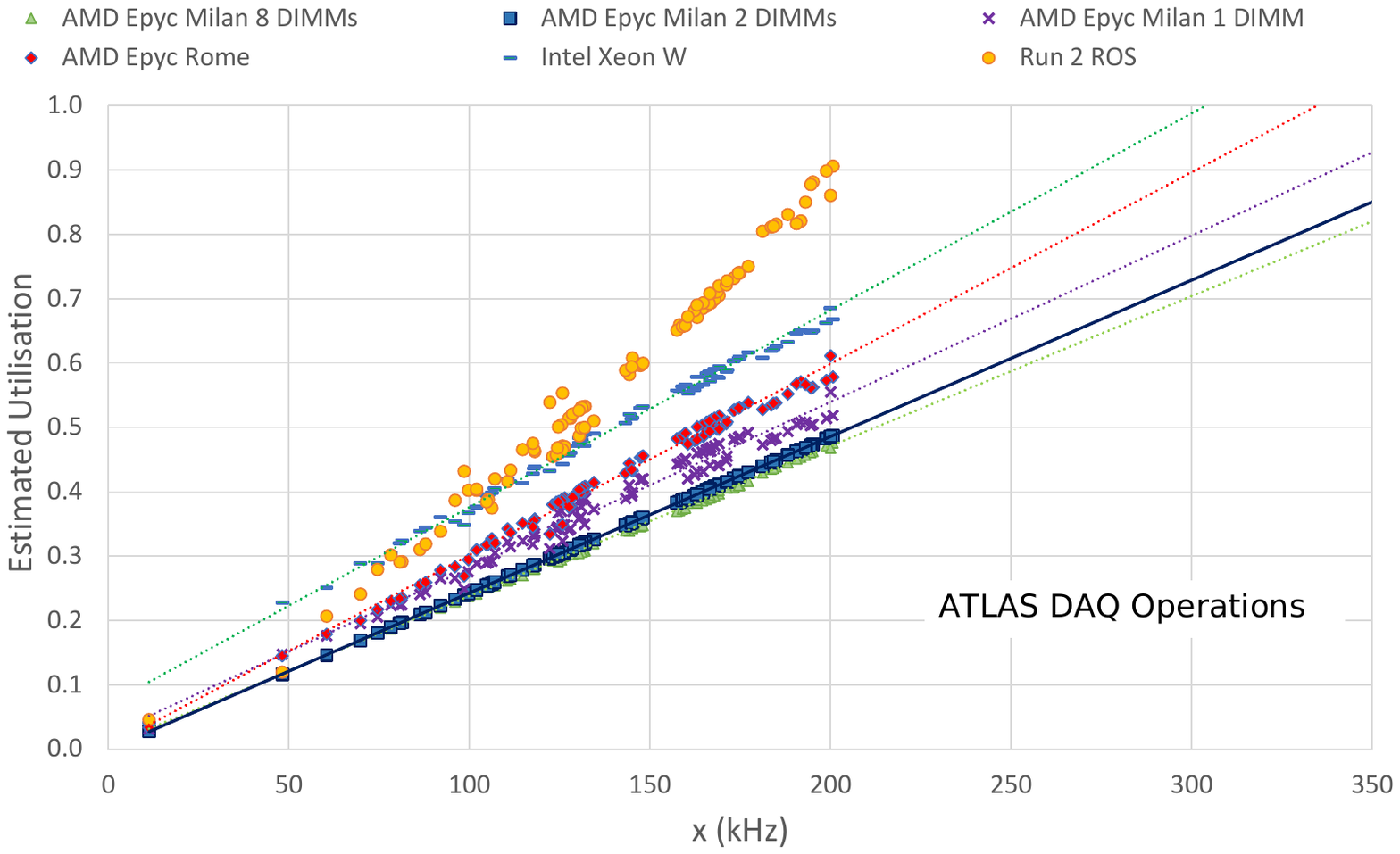}
	\caption{Utilisation values (see equation~\ref{eqn:utilisation}) for all Run~3 ROS candidate servers, along with the Run~2 server, versus the quantity $x$ (as defined equation~\ref{eqn:xdeff}).}
		\label{fig:run3utilisation}
\end{figure}

To summarise these results, Figure~\ref{fig:run3utilisation} shows the computed utilisation of all three candidates (and memory configurations), running the version of the server operating system and TDAQ software that has been deployed for the start of Run~3, compared to that of a standard ROS server from Run~2 running the operating system and version of the TDAQ software used during Run~2. The estimated utilisation is plotted against a quantity $x$, for each ROS server in the ATLAS DAQ System calculated from rates taken from a 1 hour window during a run from August 2018 (in terms of the quantities defined in Section~\ref{subsubsec:utilisation}):

\begin{equation}
x = F_{d} + 2 * F_{e} + 0.14 * F_{i} + 0.035 * \frac{S}{1000}
\label{eqn:xdeff}
\end{equation}

As can be seen from the plot, the estimated utilisations depend approximately linearly on $x$. On the basis of this quantity and with the help of the lines fitted to the data points in the plot the effect of an increase of the rates on the utilisation can be estimated.  For the Run~3 candidate (AMD Epyc Milan with 2 x 16 GB DIMMS) the linear fit is best as $x$ is in fact about equal to the utilisation (\ref{eqn:utilisation}) divided by $t_{d}$ for that configuration.
As also can be seen, the utilisation for the Run~2 ROS server, along with the other non-selected candidates/configurations, is systematically higher than the AMD Epyc Milan (2 or 8 DIMMs of 16 GB). 
Based on these results, and other functional and stability tests, the AMD Epyc Milan based server with 2 DIMMs was chosen to become the Run~3 ROS server.

\subsection{AMD Epyc `Milan' Compared to Run~3 Trigger Rate Predictions}

\begin{figure}[hbtp!]
	\centering
	\includegraphics[width=1.0\linewidth]{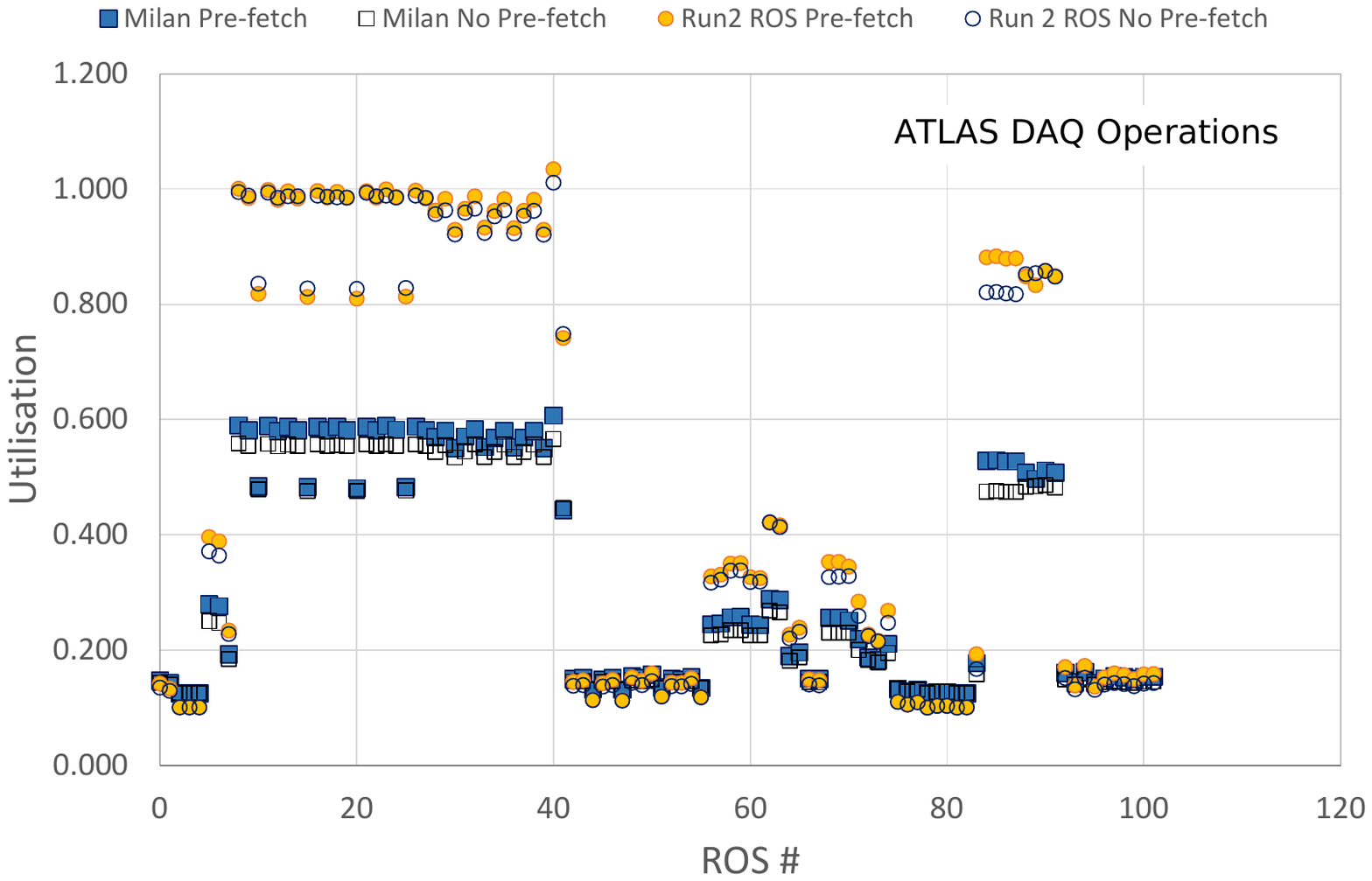}
	\caption{Comparison of estimated ROS utilisation (see equation~\ref{eqn:utilisation}) for individual servers (same data as Figure~\ref{fig:l2_saturation}), for Run~2 and Run~3 ROS servers with 32 GB RAM configuration. Note that the Run~2 server was operated using the Run~2 operating system and software conditions, whereas the Run~3 server was operated with the operating system and software conditions optimised for the start of Run~3.}
		\label{fig:milantrigger}
\end{figure}

Using the results above, it is possible to estimate how the AMD Epyc `Milan' server would perform against the expected trigger request rate requirements from Section~\ref{subsubsec:limitations}. Figure~\ref{fig:milantrigger} shows the performance of the system for both pre-fetch and no pre-fetch mode compared to the Run~2 server, where the former is running the operating system and version of the TDAQ software that has been deployed for the start of Run~3 and the latter the versions used in Run~2. The Milan server can be seen to be able to satisfy the expected rates for all subdetectors with significant operational margin, with or without pre-fetching.

\clearpage
\section{Conclusions \& Outlook}
\label{sec:conclusions}

In this paper the design and operation of the upgraded ATLAS ROS in LHC Run~2 have been presented, as well as preparations and updates for Run~3. The upgraded system makes use of new technology and concepts to significantly improve on the performance of its predecessor in Run~1. The RobinNP card, based on the ALICE C-RORC hardware, provides increased link density, memory capacity and PCIe bandwidth. By moving the majority of the buffer management and request handling logic into host software, making use of the novel FIFO duplicator concept, the RobinNP is a higher performance and lower maintenance component than the original ROBIN. The upgraded ROS performed well during Run~2, able to support request loads well beyond the original specification, also facilitating the deployment of a significantly larger HLT farm than would otherwise have been possible and thus improving the physics reach of the experiment.

During Run~3, the ROS continues to operate as the primary readout path for the majority of ATLAS systems. The new FELIX system~\cite{PanduroVazquez:2784274} reads out new detector systems and trigger electronics ahead of full scale deployment for LHC Run~4. In terms of performance requirements, the load on the ROS in Run~3 is not expected to be significantly larger than Run~2. However, some performance limits were reached at the end of Run~2, depending on the configuration of the HLT. Due to this and the age of the machines, an effort was undertaken to choose a new, higher performance, host server to take advantage of technological evolution over the past few years. The selected server is based on the new AMD Epyc 'Milan' generation of CPUs, and has been demonstrated to satisfy all expected Run~3 trigger rate requirements with significant operational margin. All ROS hosts have been replaced with this newer model, with the hosted RobinNPs otherwise unchanged. While the C-RORC hardware can in principle support double the bandwidth that is used for the RobinNP, there is no operational requirement from ATLAS to explore the necessary firmware modifications to utilise this extra capacity. Such a change would also require significant refactoring of the ATLAS dataflow network to deal with the extra data volume. As such, it is expected that no firmware modifications will be made to the RobinNP during Run~3.

With the server upgrade, plus the option to update the RobinNP firmware if needed, the ATLAS ROS is expected to satisfy all ATLAS performance requirements for Run~3 with significant margin. The replacement servers will also ensure that the excellent reliability of the system during Run~2 is maintained through to the point that the system is succeeded by FELIX ATLAS-wide in Run~4.




\newpage
\bibliographystyle{atlasnote-jv}
\bibliography{rospaper}
\end{document}